\documentclass[aps,prb,reprint,groupedaddress,twocolumn]{revtex4-2}
\usepackage{graphicx}
\usepackage{amsmath}
\usepackage{amssymb}
\usepackage{xcolor}
\usepackage{float}
\usepackage[utf8]{inputenc}

\newcommand{\reftextit}[1]{}
\newcommand{\bra}[1]{\left< #1 \right|}
\newcommand{\ket}[1]{\left| #1 \right>}

\newcommand{\br}{\mathbf{r}}
\newcommand{\bk}{\mathbf{k}}

\newcommand{\bq}{\mathbf{q}}
\newcommand{\bl}{\mathbf{l}}
\newcommand{\bp}{\mathbf{p}}
\newcommand{\bG}{\mathbf{G}}
\newcommand{\bR}{\mathbf{R}}

\newcommand{\wfc}[5]{\phi^{#1,#2}_{#3,#4}(#5)}
\newcommand{\en}[2]{\varepsilon^{#1}_{#2}}

\begin{document}

\title{Microscopic Theory of Exciton-Exciton Annihilation in Two-Dimensional Semiconductors}

\author{Alexander Steinhoff}
\affiliation{Institute for Theoretical Physics, University of Bremen, 28334 Bremen, Germany}
\email{asteinhoff@itp.uni-bremen.de}
\author{Matthias Florian}
\affiliation{Institute for Theoretical Physics, University of Bremen, 28334 Bremen, Germany}
\affiliation{Department of Electrical Engineering and Computer Science, University of Michigan, Ann Arbor, MI, USA}
\thanks{Present address}
\author{Frank Jahnke}
\affiliation{Institute for Theoretical Physics, University of Bremen, 28334 Bremen, Germany}


\begin{abstract}
Auger-like exciton-exciton annihilation (EEA) is considered the key fundamental limitation to quantum yield in devices based on excitons in two-dimensional (2d) materials.
%
Since it is challenging to experimentally disentangle EEA from competing processes, 
guidance of a quantitative theory is highly desirable. 
The very nature of EEA requires a material-realistic description 
that is not available to date.
%
We present a many-body theory of EEA based on first-principle band structures and Coulomb interaction matrix elements that goes beyond an effective bosonic picture.
Applying our theory to monolayer MoS$_2$ encapsulated in hexagonal BN, we obtain an EEA coefficient in the order of $10^{-3}$ cm$^{2}$s$^{-1}$ at room temperature,
suggesting that carrier losses are often dominated by other processes, such as defect-assisted scattering.
Our studies open a perspective to quantify the efficiency of intrinsic EEA processes in various 2d materials in the focus of modern materials research.
\end{abstract}

\maketitle

\section{Introduction}

Auger recombination is a Coulomb interaction process where an electron-hole pair recombines nonradiatively by transferring the excess energy
to another charge carrier. Since this process is operative at high carrier densities, it has been discussed for decades as a loss mechanism in optoelectronic 
devices that use highly excited semiconductors as active material \cite{sermage_comparison_1986, haug_auger_1992, fuchs_auger_1993, klimov_quantization_2000, dukovic_observation_2004}.
\\In systems with strong Coulomb interaction, bound electron-hole pairs termed excitons can dominate the dynamics, as long as the density of excited carriers is much smaller than the 
Mott density \cite{semkat_ionization_2009, steinhoff_exciton_2017}. In this situation, Auger scattering is expected to take place between two excitons
instead of three unbound particles. 
A prominent material class with ultra-strong Coulomb interaction are atomically thin transition metal dichalcogenides (TMDs).
The discussion of Auger-like exciton-exciton annihilation (EEA) in TMDs has been started by the works of Sun et al. \cite{sun_observation_2014} and Kumar et al. \cite{kumar_exciton-exciton_2014}. Since then, a range of EEA coefficients from several $10^{-3}$ cm$^{2}$s$^{-1}$ to about $0.1$ cm$^{2}$s$^{-1}$ have been found experimentally
for different TMD materials \cite{sun_observation_2014, kumar_exciton-exciton_2014, yuan_exciton_2015, mouri_nonlinear_2014, poellmann_resonant_2015, sim_role_2020, perea-causin_exciton_2019}. Many of the early experiments have been performed on a SiO$_2$ substrate. A strong reduction of exciton recombination for TMDs encapsulated in hexagonal boron nitride (hBN) \cite{hoshi_suppression_2017,cordovilla_leon_hot_2019,zipfel_exciton_2020} suggests that passivation plays a role, similar as for the quality of optical spectra \cite{cadiz_excitonic_2017}. 
The wide spectrum of experimental results and the sensitivity of exciton lifetimes to extrinsic effects call for a theoretical prediction of intrinsic EEA coefficients to quantify their 
impact on device performance.
\\There have been theory studies based on 
a two-band $k\cdot p$-model \cite{konabe_effect_2014}, 
Monte Carlo simulations 
\cite{mouri_nonlinear_2014} and 
a dipole-dipole interaction model assuming spatially localized excitons \cite{chatterjee_low-temperature_2019}. 
While these are low-energy models, it has been pointed out in Refs.~\cite{danovich_auger_2016,han_exciton_2018} 
that EEA involves target states from higher bands. In Ref.~\cite{han_exciton_2018}, the $k\cdot p$-approach was extended to a third band, 
which relies on the assumption that the target states selected by energy conservation are close to the third band's extremum.
\\ In translationally invariant systems, quantitative predictions of EEA efficiency generally require material-realistic input to determine overlaps of Bloch states in different bands beyond high-symmetry points. 
The challenge is
to combine the large phase space of target states with a theory for exciton-exciton interaction in second order of the Coulomb potential.
A first step towards a universal description of EEA processes has been taken recently for confined semiconductor nanostructures \cite{philbin_electronhole_2018, philbin_area_2020}. Here, EEA rates have been calculated in the so-called interacting framework, which includes electron-hole correlations within the initial electron-hole pairs, while neglecting those between excitons or final-state carriers.
%
\\
In this paper, we present a theory of EEA in 2d materials based on a many-body description of exciton-exciton scattering processes 
using
a general band structure and Coulomb interaction matrix elements from density functional theory (DFT).
It is shown that the resulting equations of motion (EOM) can not be obtained from a purely bosonic Hamiltonian 
with effective exciton-exciton interaction matrix elements since the fermionic substructure of excitons would be otherwise neglected.
The presented approach consistently takes into account all electron-hole correlations on a two-particle level.
We apply our theory to quantify 
EEA coefficients in monolayer MoS$_2$
and
analyze 
how different Bloch states contribute to the exciton-exciton scattering. We also study the influence of temperature and dielectric environmental screening on 
EEA, finding an inverse temperature dependence that is much stronger than the dependence on substrate dielectric constants.

\section{Theory}
\begin{figure}
\centering
\includegraphics[width=1.\columnwidth]{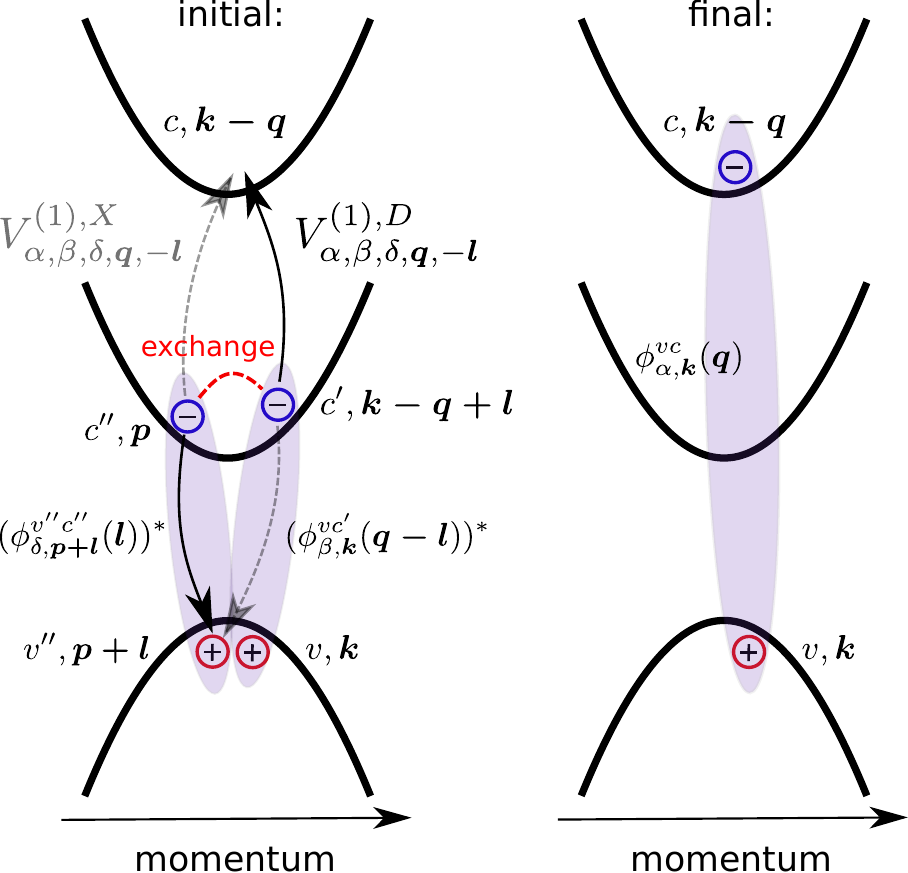}
\caption{Illustration of an electron-mediated EEA process described by effective exciton-exciton interaction matrix elements $V^{(1),D/X}_{\alpha,\beta,\delta,\bq,-\bl}$.
In the direct (D) process depicted by solid black arrows, an exciton in state $\big|\delta,\bl\big> $ is annihilated by means of inter-gap electron scattering. 
Excess momentum and energy are transferred to a second electron that is part of an exciton in state $\big|\beta,\bq-\bl\big> $, 
which is thereby excited into state $\big|\alpha,\boldsymbol{q}\big>$. In the exchange-like (X) process shown by dashed grey arrows, electrons in states 
$\big|c'',\boldsymbol{p}\big> $ and $\big|c',\boldsymbol{k-q+l}\big> $ are exchanged between the excitons, leading to the same final state.
Since Coulomb interaction conserves spin, the direct process is allowed for like-spin excitons $\big|\delta,\bl\big> $ only.
}
\label{fig:EEA_scheme}
\end{figure}

Our starting point is the Hamiltonian for Bloch electrons interacting via a statically screened Coulomb potential:
%
\begin{equation}
 \begin{split}
H&=H_0+H_{\textrm{Coul}} \\&=
\sum_{\bk,c}\varepsilon^{c}_{\bk}a^{\dagger}_{\bk,c}a^{\phantom\dagger}_{\bk,c}+\sum_{\bk,v}\varepsilon^{v}_{\bk}a^{\dagger}_{\bk,v}a^{\phantom\dagger}_{\bk,v}\\
&+\frac{1}{2\mathcal{A}}\sum_{\substack{\bk,\bk',\bq \\ \lambda,\nu,\nu',\lambda'}}V^{\lambda,\nu,\nu',\lambda'}_{\bk,\bk',\bk'+\bq,\bk-\bq} 
a^{\dagger}_{\bk,\lambda}a^{\dagger}_{\bk',\nu}a^{\phantom\dagger}_{\bk'+\bq,\nu'}a^{\phantom\dagger}_{\bk-\bq,\lambda'}\,,
\end{split}
\label{eq:H_coul}
\end{equation}
where $a^{\dagger}_{\bk,\lambda}$ and $a^{\phantom\dagger}_{\bk,\lambda}$ denote carrier creation and annihilation operators, respectively, $\varepsilon^{c/v}_{\bk}$
is the energy of a carrier with momentum $\bk$ in a conduction/valence band, $V^{\lambda,\nu,\nu',\lambda'}_{\bk,\bk',\bk'+\bq,\bk-\bq}$ are Coulomb interaction matrix elements, 
and $\mathcal{A}$ is the crystal area. EEA emerges as a higher-order carrier-carrier interaction process within the dynamics of microscopic exciton populations, which are described by two-particle correlations (doublets)
$n_{\alpha,\bq}= \Delta \big\langle X^{\dagger}_{\alpha,\bq}X^{\phantom\dagger}_{\alpha,\bq}  \big\rangle$ \cite{kira_many-body_2006}. Here, we introduced the exciton creation operator
$X^{\dagger}_{\alpha,\bq}=\sum_{\bk,v,c}\phi^{vc}_{\alpha,\bk}(\bq)a_{\bk-\bq,\textrm{c}}^{\dagger}a_{\bk,\textrm{v}}^{\phantom\dagger} $,
where $\bq$ is the total momentum of the electron-hole pair, while $\alpha$ is the quantum number belonging to the relative motion of electron and hole.
The wave functions $\phi^{vc}_{\alpha,\bk}(\bq)$ are solutions of the Bethe-Salpeter equation (BSE) in the absence of photoexcited carriers
%
\begin{equation}
 \begin{split}
&(\varepsilon^{\textrm{c}}_{\bk-\bq}-\varepsilon^{\textrm{v}}_{\bk}-E_{\alpha,\bq})\phi^{vc}_{\alpha,\bk}(\bq) \\
-&\frac{1}{\mathcal{A}}\sum_{\bk',v',c'}
V^{c,v',v,c'}_{\bk-\bq,\bk',\bk,\bk'-\bq}\phi^{v'c'}_{\alpha,\bk'}(\bq)=0\,,
\end{split}
\label{eq:wannier}
\end{equation}
which comprises bound exciton states as well as unbound scattering states. Here, $E_{\alpha,\bq}$ are the two-particle eigenenergies.
We
assume that the dynamics is governed by correlated electron-hole pairs, so that we can formulate a closed system of equations for the doublets $n_{\alpha,\bq}$, while single-particle occupancies 
$f^{\lambda}_{\bk}=\Delta\big\langle a^{\dagger}_{\bk,\lambda}a^{\phantom\dagger}_{\bk,\lambda}  \big\rangle$ are dropped.
This is justified in an intermediate density regime well below the so-called Mott density, that marks a transition to a quantum phase of unbound electrons and holes \cite{semkat_ionization_2009, steinhoff_exciton_2017}, or at low density after resonant optical excitation of bound states \cite{selig_dark_2018}.
Since excitons are globally charge neutral, excitation-induced many-body renormalizations of the BSE eigenstates and eigenenergies can be neglected. A discussion of excitation-induced effects is given in Ref.~\cite{schleife_optical_2011}.
\\
The EOM for the
$n_{\alpha,\bq}$ including EEA 
are derived by applying the
cluster expansion technique \cite{kira_many-body_2006}. 
For details, we refer to the Supporting Information (SI).
The resulting hierarchy of equations
is truncated 
%
by introducing a phenomenological damping $\Gamma$ of three-particle correlations
and using the Markov approximation, which yields:
\begin{equation}
 \begin{split}
\frac{d}{dt} n_{\alpha,\bq}\big|_{\textrm{EEA}} = \frac{1}{\mathcal{A}^2}\sum_{\bl}\sum_{\beta\delta}\Big[&\Xi^{(\alpha,\bq)\leftrightarrow(\beta,\bq-\bl),(\delta,\bl)} \\
 - &\Xi^{(\beta,\bq+\bl)\leftrightarrow(\alpha,\bq),(\delta,\bl)} \Big]
\end{split}
\label{eq:eom_nqnu}
\end{equation}
with the scattering rates
\begin{equation}
 \begin{split}
&\Xi^{(\alpha,\bq)\leftrightarrow(\beta,\bq-\bl),(\delta,\bl)}= 
  \\
 &\frac{2}{\hbar}\,\textrm{Im}\Bigg\{
\frac{V^{D,\alpha,\beta,\delta}_{\bq,-\bl}}{E_{\alpha,\bq}-E_{\beta,\bq-\bl} - E_{\delta,\bl} -i\Gamma }\times\\
\Big\{&(V^{\alpha,\beta,\delta}_{\bq,-\bl})^*
 (n_{\beta,\bq-\bl}n_{\delta,\bl} -n_{\alpha,\bq}( n_{\delta,\bl}+1)) \\
 + &(V^{\alpha,\delta,\beta}_{\bq,-\bq+\bl})^*
 (n_{\beta,\bq-\bl} n_{\delta,\bl} - n_{\alpha,\bq}(n_{\beta,\bq-\bl} +1))
 \Big\} \Bigg\} \,, \\ 
 &\Xi^{(\beta,\bq+\bl)\leftrightarrow(\alpha,\bq),(\delta,\bl)}=
 \\
 &\frac{2}{\hbar}\,\textrm{Im}\Bigg\{
 \frac{V^{\beta,\alpha,\delta}_{\bq+\bl,-\bl}}{E_{\beta,\bq+\bl}-E_{\alpha,\bq} - E_{\delta,\bl} -i\Gamma }\times\\
\Big\{&(V^{\beta,\alpha,\delta}_{\bq+\bl,-\bl})^*   
 (n_{\alpha,\bq}n_{\delta,\bl} -n_{\beta,\bq+\bl}( n_{\delta,\bl}+1)) \\
 +& (V^{\beta,\delta,\alpha}_{\bq+\bl,-\bq})^*
 (n_{\alpha,\bq} n_{\delta,\bl}-n_{\beta,\bq+\bl}(n_{\alpha,\bq}+1))
 \Big\}
 \Bigg\}\,.
\end{split}
\label{eq:eom_nqnu2}
\end{equation}
%
EEA is a second-order process in terms of effective exciton-exciton interaction matrix elements
\begin{equation}
\begin{split}
V^{\alpha,\beta,\delta}_{\bq,-\bl}&=V^{D,\alpha,\beta,\delta}_{\bq,-\bl}-V^{X,\alpha,\beta,\delta}_{\bq,-\bl},\\
V^{D/X,\alpha,\beta,\delta}_{\bq,-\bl}&=V^{(1),D/X}_{\alpha,\beta,\delta,\bq,-\bl} - 
 V^{(2),D/X}_{\alpha,\beta,\delta,\bq-\bl,-\bl}\,, \\
V^{(1),D}_{\alpha,\beta,\delta,\bq,-\bl}&= \\ \sum_{\bk,\bp}\sum_{v,c,c',v'',c''} 
&\wfc{v}{c}{\alpha}{\bk}{\bq}(\wfc{v''}{c''}{\delta}{\bp+\bl}{\bl})^*(\wfc{v}{c'}{\beta}{\bk}{\bq-\bl})^*\times\\
&
 V^{c',c'',v'',c}_{\bk-\bq+\bl,\bp,\bp+\bl,\bk-\bq}\,, \\
 V^{(1),X}_{\alpha,\beta,\delta,\bq,-\bl}&= \\ \sum_{\bk,\bp}\sum_{v,c,c',v'',c''} 
&\wfc{v}{c}{\alpha}{\bk}{\bq}(\wfc{v''}{c''}{\delta}{\bp+\bl}{\bl})^*(\wfc{v}{c'}{\beta}{\bk}{\bq-\bl})^*\times\\
&
 V^{c'',c',v'',c}_{\bp,\bk-\bq+\bl,\bp+\bl,\bk-\bq}\,, \\
V^{(2),D}_{\alpha,\beta,\delta,\bq,-\bl}&= \\ \sum_{\bk,\bp}\sum_{v,c,v',v'',c''} 
&(\wfc{v}{c}{\beta}{\bk}{\bq})^*(\wfc{v''}{c''}{\delta}{\bp+\bl}{\bl})^*\wfc{v'}{c}{\alpha}{\bk+\bl}{\bq+\bl}\times\\
&
V^{v',c'',v'',v}_{\bk+\bl,\bp,\bp+\bl,\bk}\,,\\
V^{(2),X}_{\alpha,\beta,\delta,\bq,-\bl}&= \\ \sum_{\bk,\bp}\sum_{v,c,v',v'',c''} 
&(\wfc{v}{c}{\beta}{\bk}{\bq})^*(\wfc{v''}{c''}{\delta}{\bp+\bl}{\bl})^*\wfc{v'}{c}{\alpha}{\bk+\bl}{\bq+\bl}\times\\
&
V^{c'',v',v'',v}_{\bp,\bk+\bl,\bp+\bl,\bk}\,,
\end{split}
\label{eq:coul_me}
\end{equation}
where $V^{(1)}$ and $V^{(2)}$ describe Auger-like scattering of electrons and holes, respectively. 
Hence the effective interaction is composed of elementary scattering processes weighted by the respective two-particle wave functions, where
the relative minus sign between $V^{(1)}$ and $V^{(2)}$ reflects the opposite charges of electrons and holes. This is similar in exciton-phonon interaction
\cite{selig_excitonic_2016}.
An elementary electron scattering process belonging to the rate $\Xi^{(\alpha,\bq)\leftrightarrow(\beta,\bq-\bl),(\delta,\bl)}$ is schematically shown in Fig.~\ref{fig:EEA_scheme}. 
Energy conservation is softened by the finite exciton lifetimes.
\\Note that the EOM do not have the form of a Boltzmann equation. 
Accordingly, as we demonstrate in the SI, the equations can not be derived from an exciton Hamiltonian with effective
exciton-exciton interaction and bosonic commutation relations. This is due to the fact that excitons are composite particles with a 
fermionic substructure that can not be captured in a purely bosonic picture. 
In our theory, additional scattering terms emerge from the exchange of electrons or holes between two excitons, as symbolized by the dashed red line 
in Fig.~\ref{fig:EEA_scheme}. On the other hand, a bosonic theory can only account for exchange of entire excitons.
It has been pointed out in more general terms by M. Combescot et al. that it is not possible to formulate a closed expression for an 
effective exciton-exciton interaction potential \cite{combescot_effective_2002}.
Although we have introduced exciton-exciton interaction
matrix elements (\ref{eq:coul_me}), these can not be interpreted as such an effective interaction potential, since the very structure of the
EOM (\ref{eq:eom_nqnu}) is beyond a purely bosonic picture. An alternative approach to treat excitons as non-ideal bosons is to introduce fermionic corrections 
into the commutator of exciton operators \cite{katsch_theory_2018}.
As we show in the Appendix, our theory can only be mapped to an effective bosonic Hamiltonian for an exciton distribution close to equilibrium if the exchange of whole excitons is neglected. If fermionic exchange effects are included in the effective matrix elements, EEA efficiency in encapsulated MoS$_2$ is overestimated by a factor $2$. This is partly remedied by neglecting fermionic exchange as well due to a compensation between the different exchange effects.
\\The EOM are completed by a phenomenological exciton-phonon scattering term in relaxation-time approximation that accounts for relaxation and cooling
of the exciton gas:
\begin{equation}
\begin{split}
\frac{d}{dt} n_{\alpha,\bq}\big|_{\textrm{relax}}= \frac{N_{\alpha,\bq}(T) - n_{\alpha,\bq}}{\tau_{\textrm{relax}}}
\end{split}
\label{eq:relax}
\end{equation}
with Bose functions $N_{\alpha,\bq}(T)$. We choose $\tau_{\textrm{relax}}=50$ fs as relaxation time \cite{selig_excitonic_2016}. Since exciton-phonon scattering is
several orders of magnitude faster than EEA, 
we assume that a microscopic description of exciton-phonon coupling would not improve our results.
\\We combine our theory of EEA with band structures and screened Coulomb matrix elements on a DFT level as input for Eqs.~(\ref{eq:wannier}), (\ref{eq:coul_me}) and (\ref{eq:eom_nqnu}). 
Details on the DFT calculation are provided in the SI. The main effect of a GW correction would be an increased band gap, which is reduced again due to environmental screening \cite{thygesen_calculating_2017}. We mimic this effect by artificially increasing the band gap such that the bright exciton energy is 
$E_{\textrm{1s,bright}}=1900$ meV \cite{cadiz_excitonic_2017}. 
As an interface between first-principle and excited-carrier theory we utilize a lattice Hamiltonian formulated in a localized basis of Wannier orbitals (d$_{z^2}$, d$_{xz}$, d$_{yz}$, d$_{x^2-y^2}$ and d$_{xy}$ for Mo, p$_x$, p$_y$ and p$_z$ for S). Spin-orbit interaction is included using an on-site $\boldsymbol{L\cdot S}$-coupling Hamiltonian.
From the band structure obtained by diagonalization of the lattice Hamiltonian we consider two valence bands and three conduction bands for each spin degree of freedom, which is sufficient to capture all scattering states at about twice the 1s-exciton energy involved in the annihilation of 1s-excitons. Coulomb matrix elements including environmental screening effects are parametrized as a function of $|\bq|$ using the Wannier function continuum electrostatics approach \cite{rosner_wannier_2015} in the localized basis. Unless stated otherwise, we assume a dielectric environment given by hBN encapsulation layers with a dielectric constant of $\varepsilon_{\textrm{hBN}} = \sqrt{4.95 \cdot 2.86}$ \cite{artus_natural_2018}. In addition a narrow gap of $0.3$ nm between the monolayer and the surrounding hBN layers has been taken into account \cite{florian_dielectric_2018}. 
\\The numerical simulation of EEA is constrained by the high-dimensionality of the problem. In the following, we focus on Bloch states in the 
K-valley, where bright excitons are located in monolayer MoS$_2$. An explicit treatment of excitons in the equivalent K'-valley is not necessary assuming that Auger-like 
EEA involving carrier scattering between K- and K'-valley is inefficient due to the large momentum transfer it involves.
Numerical convergence of the results is discussed in the SI.
\section{Results}
\begin{figure}
\centering
\includegraphics[width=\columnwidth]{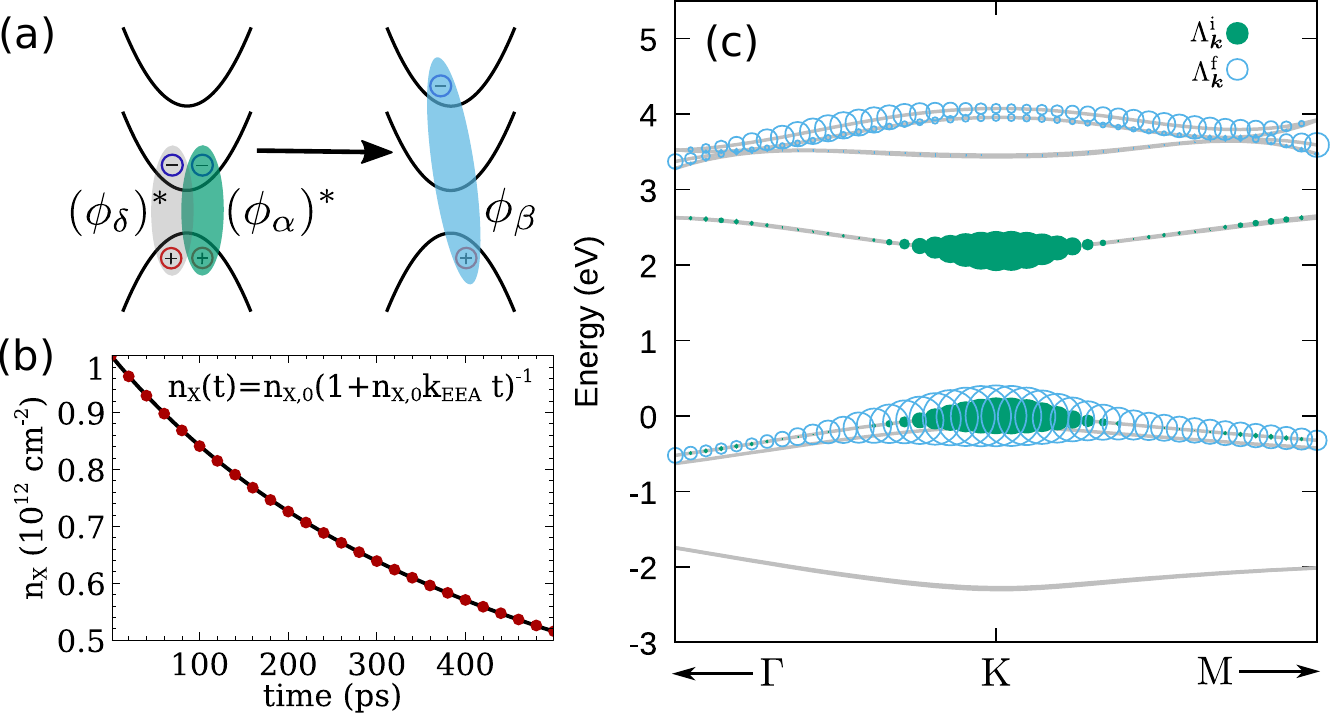}
\caption{
(a) Fundamental EEA process: An exciton in state $\ket{\delta}$ is annihilated, while a second exciton is promoted from low-energy 
state $\ket{\alpha}$ to high-energy state $\ket{\beta}$.
(b) Time dependence of the total exciton density $n_{\textrm{X}}$ in hBN-encapsulated MoS$_2$ at $T=300$ K. Red dots correspond to the 
numerical solution of the EOM (\ref{eq:eom_nqnu}) and (\ref{eq:relax}), while the analytic solution of the simplified ODE (\ref{eq:simple_ODE}) is shown as a black solid line. 
(c) Contribution of Bloch bands to EEA scattering rates.
The symbol size represents the projection of initial (full green circles) and final (open blue circles) two-particle states onto Bloch states
according to the corresponding two-particle wave functions, as described by the 
quantities $\Lambda^{\textrm{i}}_{\bk}$ and
$\Lambda^{\textrm{f}}_{\bk}$ defined in Eq.~(\ref{eq:DOS}). The single-particle band structure is shown in grey.
}
\label{fig:key_results}
\end{figure}

\begin{figure}
\centering
\includegraphics[width=\columnwidth]{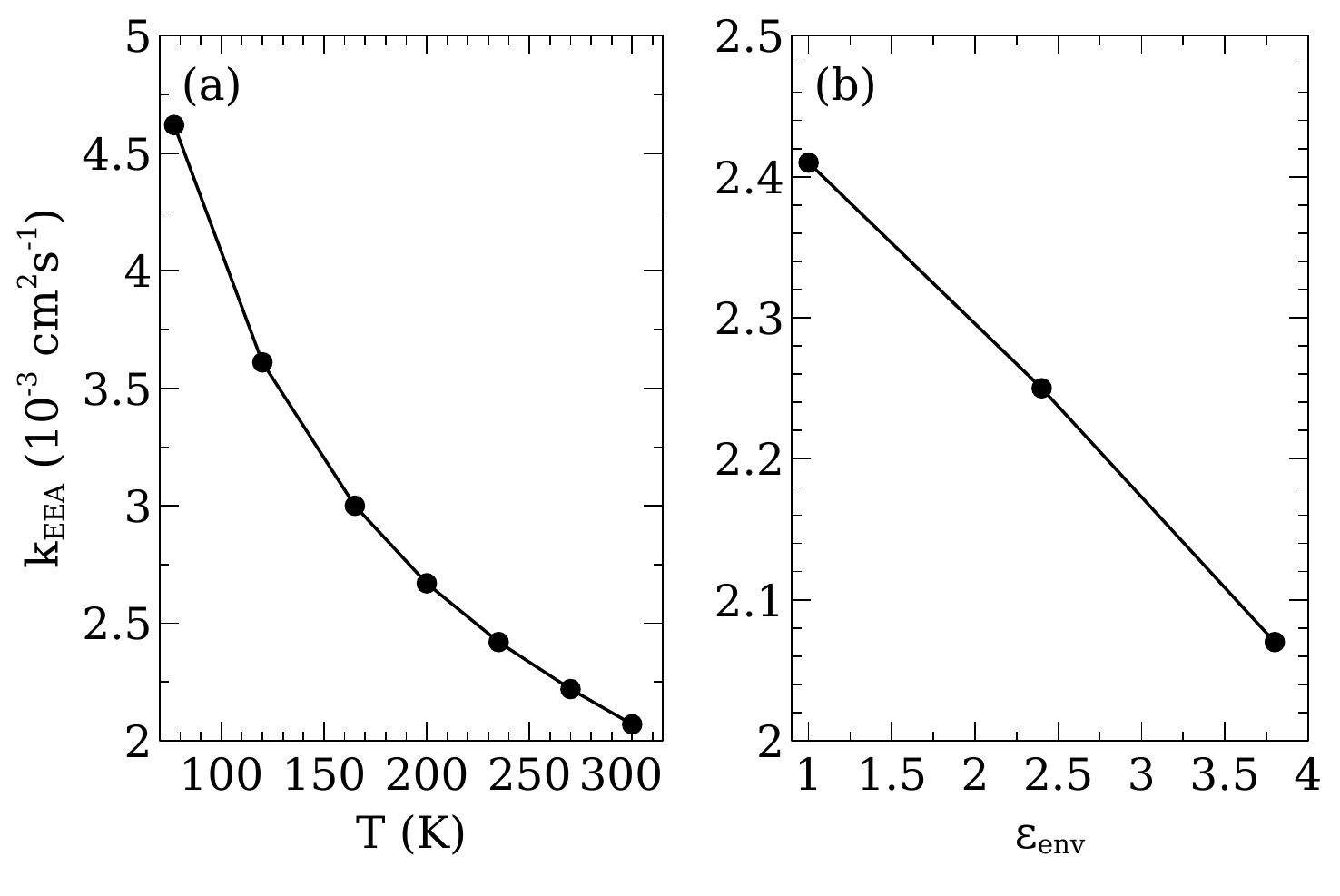}
\caption{(a) Temperature dependence of EEA coefficient in hBN-encapsulated MoS$_2$. 
(b) Dependence of EEA coefficient for MoS$_2$ on average dielectric constant of environment 
$\varepsilon_{\textrm{env}}=(\varepsilon_{\textrm{top}}+\varepsilon_{\textrm{bottom}}) / 2$ at $T=300$ K.}
\label{fig:T_eps_dep}
\end{figure}

As a first step, we diagonalize the BSE (\ref{eq:wannier}) to obtain the two-particle spectrum and wave functions, finding a bright 1s-exciton binding energy of $287$ meV.
The EOM for exciton populations are solved including EEA (\ref{eq:eom_nqnu}) and relaxation (\ref{eq:relax}) contributions. To this end, we assume that 
an incoherent exciton gas with a density of $10^{12}$ cm$^{-2}$ has formed from optically excited electron-hole pairs due to ultrafast exciton-phonon interaction \cite{selig_dark_2018} before EEA sets in.
A phenomenological damping $\Gamma=50$ meV is used. 
As we show in the SI, the dependence of our results on $\Gamma$ is weak. 
The efficiency of EEA is quantified by analyzing the time dependence of the total exciton density $n_{\textrm{X}}=\frac{1}{\mathcal{A}}\sum_{\alpha,\bq}n_{\alpha,\bq}$ as shown in 
Fig.~\ref{fig:key_results}(b). 
Since relaxation of excitons is much faster than EEA, the exciton distribution is close to equilibrium at all times. The overall behavior of 
the exciton density is therefore captured very well by a macroscopic differential equation that has been used before to discuss EEA qualitatively \cite{sun_observation_2014}: 
\begin{equation}
\begin{split}
\frac{d}{dt} n_{\textrm{X}}= -k_{\textrm{EEA}} n_{\textrm{X}}^2\,,
\end{split}
\label{eq:simple_ODE}
\end{equation}
with the solution $n_{\textrm{X}}(t)= n_{\textrm{X},0}(1+n_{\textrm{X},0}k_{\textrm{EEA}}t)^{-1}$. 
We use the solution as a fit formula to extract the EEA coefficient 
$k_{\textrm{EEA}}$ that can be conveniently compared to experiment. 
The numerically converged value is estimated as 
$k_{\textrm{EEA}}=2.8\times10^{-3}$ cm$^{2}$s$^{-1}$.
\\In Fig.~\ref{fig:key_results}(c), we analyze the microscopic contribution of different Bloch states to the EEA dynamics in terms of the scattering rates between
two-particle states $\ket{\alpha,\bq}$, $\ket{\delta,\bl}$ on one side and $\ket{\beta,\bq+\bl}$ on the other side at time $t=0$ as defined in Eq.~(\ref{eq:eom_nqnu}). 
The
dominant scattering channel involves the 1s-exciton states $\ket{\alpha,\bq}$, $\ket{\delta,\bl}$ as
initial states and high-energy states $\ket{\beta,\bq+\bl}$ as final states.
As a benchmark, we compute the weighted quantities
\begin{equation}
\begin{split}
\Lambda^{\textrm{i},c}_{\bk}&=\sum_{\alpha,\beta,\delta,\bl,\bq',v}|\wfc{v}{c}{\alpha}{\bk+\bq'}{\bq'}|^2\Xi^{(\beta,\bq'+\bl)\leftrightarrow(\alpha,\bq'),(\delta,\bl)}\,, \\
\Lambda^{\textrm{i},v}_{\bk}&=\sum_{\alpha,\beta,\delta,\bl,\bq',c}|\wfc{v}{c}{\alpha}{\bk}{\bq'}|^2\Xi^{(\beta,\bq'+\bl)\leftrightarrow(\alpha,\bq'),(\delta,\bl)}\,, \\
\Lambda^{\textrm{f},c}_{\bk}&=\sum_{\alpha,\beta,\delta,\bq,\bq',v}|\wfc{v}{c}{\beta}{\bk+\bq'}{\bq'}|^2\Xi^{(\beta,\bq')\leftrightarrow(\alpha,\bq),(\delta,\bq'-\bq)}\,, \\
\Lambda^{\textrm{f},v}_{\bk}&=\sum_{\alpha,\beta,\delta,\bq,\bq',c}|\wfc{v}{c}{\beta}{\bk}{\bq'}|^2\Xi^{(\beta,\bq')\leftrightarrow(\alpha,\bq),(\delta,\bq'-\bq)}\,
\end{split}
\label{eq:DOS}
\end{equation}
corresponding to the contributions of conduction- and valence-band states to the initial (i) and final (f) states. 
The rates $\Xi$ microscopically determine the scattering efficiency out of state $\ket{\alpha,\bq'}$ and into state $\ket{\beta,\bq'}$, respectively, according to population factors, energy conservation and Coulomb matrix elements. The modulus square of two-particle wave functions is then used heuristically to project the rates onto the single-particle Bloch states.
Contributions to the initial state are concentrated around the fundamental band gap. Remarkably, all contributions to the final state stem from the topmost 
valence band and the fifth and sixth conduction bands, which belong to the subspace of bands with d$_{m=\pm 2}$-orbital character. This is consistent with
the discussion in Ref.~\cite{danovich_auger_2016, han_exciton_2018}.
Hence, EEA in monolayer MoS$_2$ is essentially driven by electron-assisted processes. Moreover, we find that final electron states apart from the K-point are favored, 
which justifies our material-realistic approach.
\\Fig.~\ref{fig:T_eps_dep} shows the dependence of EEA in monolayer MoS$_2$ on temperature and dielectric environmental screening.
We find that $k_{\textrm{EEA}}$ is inversely proportional to temperature. The only effect of temperature in our theory is the width of quasi-thermal Bose functions to which exciton distributions relax by virtue of the phenomenological term (\ref{eq:relax}).
We therefore deduce that an increased population of exciton states with small momentum at low temperatures is favorable for EEA.
As Fig.~\ref{fig:T_eps_dep}(b) shows, the efficiency of EEA decreases with increasing dielectric constant of the environment, which is expected due to the Coulomb nature of Auger scattering. However, the dependence is rather weak given the quadratic dependence of scattering rates on the screened Coulomb interaction. We attribute this to Auger-like EEA
relying to a large degree on scattering processes with large momentum transfer as underlined by Fig.~\ref{fig:key_results}(c). Environmental screening is most efficient at 
small momenta, while screening at large momenta is determined by the polarization of the TMD monolayer itself \cite{rosner_wannier_2015}.
Our results are in line with the weak dependence on the dielectric environmental reported in Ref.~\cite{yu_fundamental_2016}, while a stronger dependence has been found in Ref.~\cite{goodman_substrate-dependent_2020}. We infer that surface chemistry, not dielectric screening, is the decisive factor.
\\In various experiments, EEA coefficients have been extracted from time-resolved photoluminescence (PL) measurements. 
For a comparison between experiment and theory, one has to keep in mind that we limited ourselves to EEA in a single valley.
Inter-valley excitons consisting of electrons and holes from different valleys are approximately degenerate to intra-valley excitons.
Thus, if exciton relaxation is much faster than EEA, half of the exciton density occupies inter-valley states, which can only decay assisted by slow inter-valley scattering
due to momentum conservation. Our calculation therefore overestimates the fraction of excitons decaying via EEA by a factor $2$ (corresponding to state $\ket{\delta}$ in Fig.~\ref{fig:key_results}(a)), while the assisting scattering process can be provided by all excitons ($\ket{\alpha}$ in Fig.~\ref{fig:key_results}(a)). Hence, according to
Eq.~(\ref{eq:simple_ODE}), the coefficients we calculate have to be divided by $2$ to be compared to experiment, which yields $k_{\textrm{EEA}}=1.4\times10^{-3}$ cm$^{2}$s$^{-1}$ including k-mesh
convergence. 
%
\\The EEA coefficient we find is smaller than those reported in literature. The smallest coefficients $k_{\textrm{EEA}}=6\times10^{-3}$ cm$^{2}$s$^{-1}$ \cite{hoshi_suppression_2017} and $k_{\textrm{EEA}}=4\times10^{-3}$ cm$^{2}$s$^{-1}$ \cite{zipfel_exciton_2020}
for WS$_2$ encapsulated in hBN are in the same order of magnitude, while the coefficient drastically increases when the WS$_2$ is in direct contact
with a SiO$_2$ substrate. 
Since SiO$_2$-supported TMD layers show strong defect activity e.g. in PL \cite{cadiz_excitonic_2017} and STM \cite{klein_impact_2019},
it is plausible that
defect-assisted processes dominate the dynamics in such samples. 
This is corroborated by the fact that luminescence efficiency of monolayer MoS$_2$ can be drastically increased by a chemical treatment eliminating defect-mediated nonradiative recombination \cite{amani_near-unity_2015}.
Furthermore, it has been shown that defect-assisted exciton annihiliation has a contribution that
is quadratic in exciton density \cite{wang_fast_2015}, just as EEA.

There is a number of intrinsic processes that provide additional channels for exciton decay, which
implies that our theory sets a lower limit to observable EEA coefficients. We have already discussed that inter-valley scattering would provide decay channels for inter-valley
excitons. Also, phonon-assisted processes are expected to foster annihilation of excitons with large momentum \cite{danovich_auger_2016}. Radiative decay of uncorrelated
electron-hole pairs depends quadratically on carrier density \cite{kira_quantum_1999}. Since a certain fraction of carriers is always ionized, these processes can distort the EEA coefficient.
\\We finally discuss possible variations of EEA coefficients due to uncertainties of the underlying first-principle calculation. In the SI, we show how the coefficient
depends on $E_{\textrm{1s,bright}}$, which determines the final-state DOS at about $2 E_{\textrm{1s,bright}}$ ($\ket{\beta}$ in \ref{fig:key_results}(a)). 
For example, higher energies of the third conduction band can be simulated by a lower exciton energy. 
We find that a variation of $\pm 50$ meV in $E_{\textrm{1s,bright}}$ changes the results by $\pm 30\%$. Moreover, the used first-principle method
has an effect on the position of side valleys relative to the K-valley \cite{shi_quasiparticle_2013}. The influence of relative valley positions has been studied
by comparing TMD multilayer structures. EEA efficiency decreases with increasing layer number in materials where indirect excitons become dominant \cite{yuan_exciton_2015},
while the trend is weaker as long as the material remains direct \cite{sim_role_2020}. We therefore expect that intrinsic EEA is less efficient in tungsten-based TMD monolayers
than in MoS$_2$.

\section{Conclusion}
In conclusion, our microscopic theory of EEA allows, for the first time, to quantify the efficiency of intrinsic EEA, which is often masked by extrinsic processes in
experiment. The result of $k_{\textrm{EEA}}\approx10^{-3}$ cm$^{2}$s$^{-1}$ for hBN-encapsulated monolayer MoS$_2$ is consistent with observation.
As we have shown, the physics of EEA can not be captured by an effective bosonic theory.
In the future, the theory can be applied to various members of the expanding family of 2d materials, 
such as perovskites~\cite{shen_unexpectedly_2018}, black phosphorus~\cite{surrente_onset_2016,pareek_ultrafast_2020}
and TMD hetero-bilayers hosting spatially indirect excitons of dipolar nature \cite{rivera_observation_2015, tran_evidence_2019}.

\textbf{Acknowledgement}

We acknowledge financial support from the Deutsche Forschungsgemeinschaft (RTG 2247 "Quantum Mechanical Materials Modelling") as well as resources for computational time at the HLRN (Hannover/Berlin). M.F. acknowledges support by the Alexander von Humboldt foundation. The authors would also like to thank Prof. M. Kira for valuable discussions.

%


\newpage

\section{Appendix}

\renewcommand\thefigure{S\arabic{figure}}
\setcounter{figure}{0}
\renewcommand\theequation{S\arabic{equation}}
\setcounter{equation}{0}

\subsection{Density functional theory calculations, spin-orbit coupling and Coulomb matrix elements}

\begin{figure}
\centering
\includegraphics[width=\columnwidth]{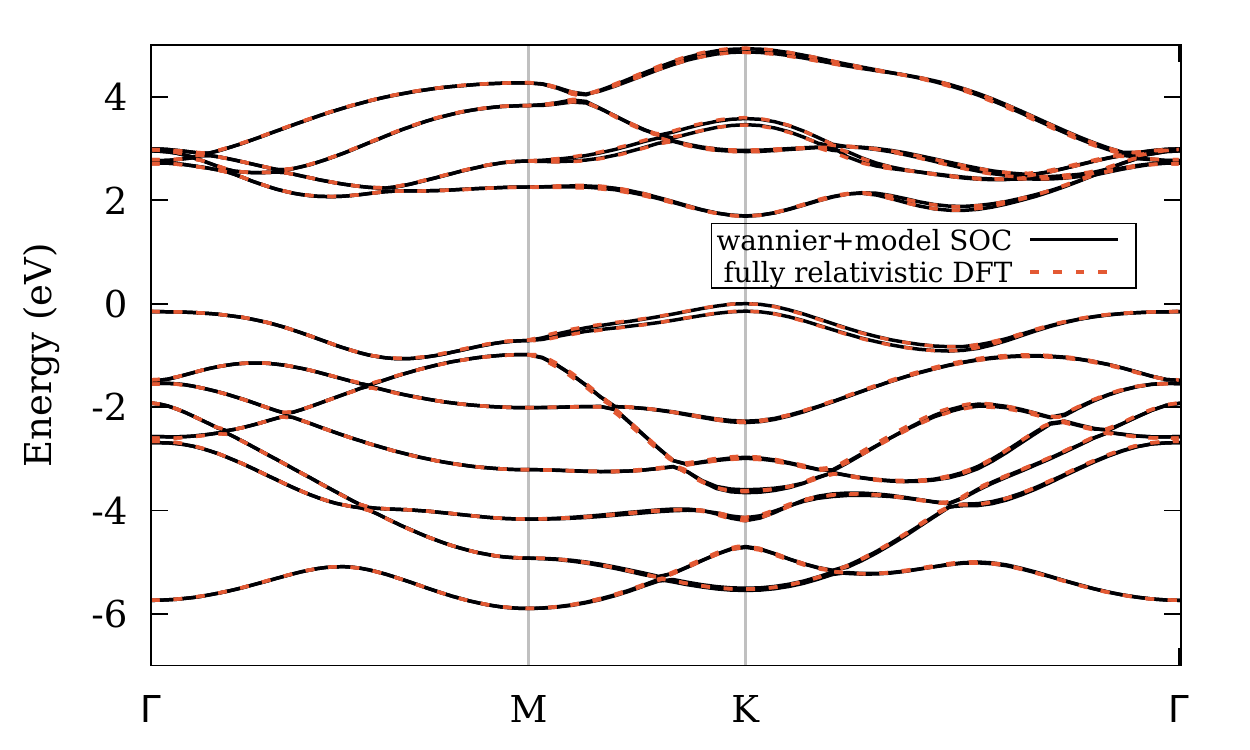}
\caption{Band structure of freestanding monolayer MoS$_2$ as obtained from a fully relativistic DFT calculation (dashed red lines) compared to a non-relativistic Wannier construction augmented by a $\boldsymbol{L\cdot S}$-Hamiltonian (solid black lines).}
\label{fig:DFT_bands}
\end{figure}

\begin{figure}
\centering
\includegraphics[width=\columnwidth]{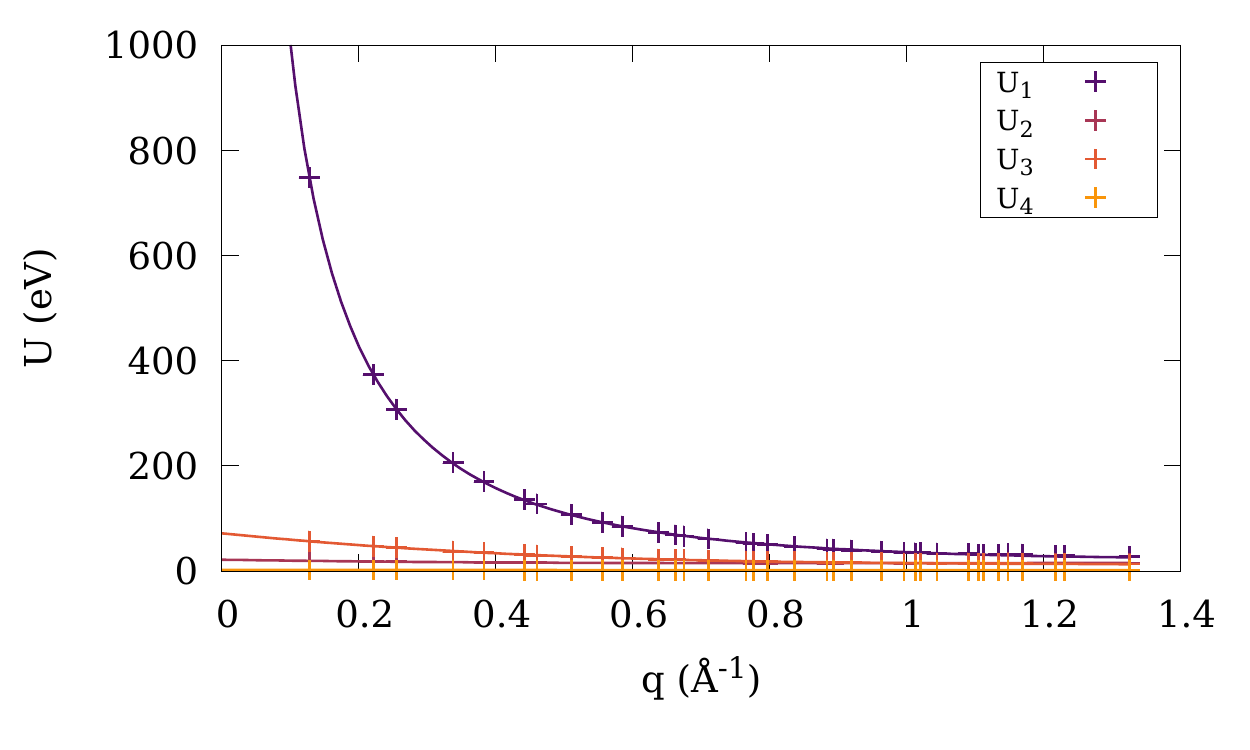}
\caption{Leading four eigenvalues of the bare Coulomb matrix for freestanding monolayer MoS$_2$ (symbols) and analytic fit functions (solid lines) as discussed in the text.}
\label{fig:U_fit}
\end{figure}

\begin{figure}
\centering
\includegraphics[width=\columnwidth]{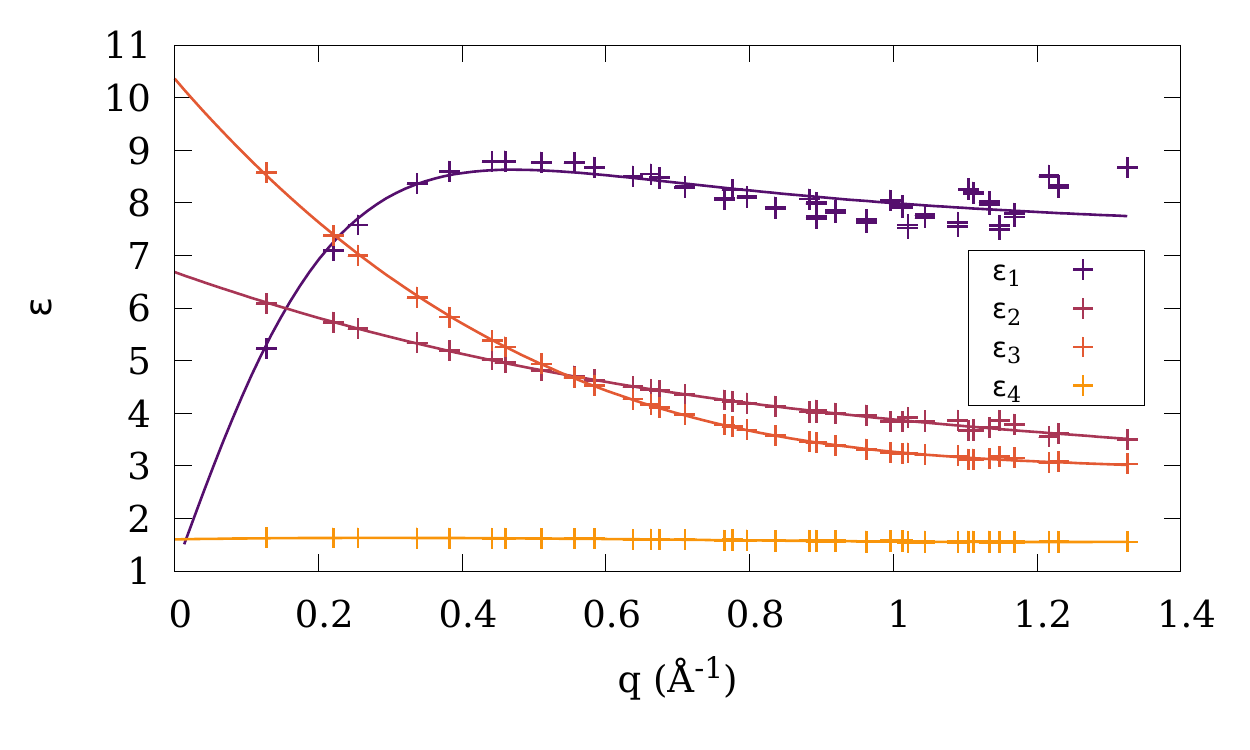}
\caption{Leading four eigenvalues of the dielectric matrix for freestanding monolayer MoS$_2$ (symbols) and analytic fit functions (solid lines) as discussed in the text.}
\label{fig:eps_fit}
\end{figure}

Density functional theory (DFT) calculations for freestanding monolayer MoS$_2$ are carried out using QUANTUM ESPRESSO V.6.6 \cite{giannozzi_quantum_2009, giannozzi_advanced_2017}. We apply the 
generalized gradient approximation (GGA) by Perdew, Burke, and Ernzerhof (PBE) \cite{perdew_generalized_1996, perdew_generalized_1997}
and use projector-augmented wave (PAW) pseudopotentials from the PSLibrary \cite{dal_corso_pseudopotentials_2014} at a plane-wave cutoff of $50$~Ry. 
Uniform meshes (including the $\Gamma$-point) with $18\times18\times1$ k-points are combined with a Fermi-Dirac smearing of $5$~mRy. 
Using a fixed lattice constant of $a=3.16$~\AA \cite{bjorkman_testing_2014, molina-sanchez_vibrational_2015} and a fixed cell height of $45$~\AA, 
forces are minimized below $10^{-3}$~eV/\AA.
\\We use RESPACK \cite{nakamura_respack:_2021} to construct a lattice Hamiltonian $H_0(\bk)$ in an 11-dimensional localized basis of Wannier orbitals (d$_{z^2}$, d$_{xz}$, d$_{yz}$, d$_{x^2-y^2}$ and d$_{xy}$ for Mo, p$_x$, p$_y$ and p$_z$ for S) from the DFT results. 
We also calculate the dielectric function as well as bare and screened Coulomb matrix elements in the localized basis. For the polarization function, a cutoff energy of $5$~Ry, $96$~bands as well as $70$~frequency points on a logarithmic grid are taken into account.
\\Spin-orbit interaction is included using an on-site $\boldsymbol{L\cdot S}$-coupling Hamiltonian along the lines of \cite{liu_three-band_2013}, which is added to the 
non-relativistic Wannier Hamiltonian:
\begin{equation}
\begin{split}
H(\bk)=I_2 \otimes H_0(\bk)+H_{\textrm{SOC}}\,.
\end{split}
\label{eq:H_tot}
\end{equation}
Here, $I_2$ is the $2\times2$ identity matrix in the Hilbert space spanned by eigenstates $\ket{\uparrow}$ and $\ket{\downarrow}$ of the spin z component (perpendicular to the monolayer). 
By treating spatial degrees of freedom and spin separately, we reduce the size of the Wannier Hamiltonian and
thereby the Coulomb matrix. We assume that the Coulomb matrix is spin-independent.
Diagonalization of $H(\bk)$ yields the band structure $\varepsilon_{\bk}^{\lambda}$ and the Bloch states $\ket{\psi_{\bk}^{\lambda}}=\sum_{\alpha} c^{\lambda}_{\alpha,\bk}\ket{\bk,\alpha} $,
where the coefficients $c^{\lambda}_{\alpha,\bk}$ describe the momentum-dependent contribution of the orbital $\alpha$ to the Bloch band $\lambda$.
The Bloch sums $\ket{\bk,\alpha}$ are connected to the localized basis via $\ket{\bk,\alpha}=\frac{1}{\sqrt{N}}\sum_{\bR}e^{i\bk\cdot\bR}\ket{\bR,\alpha}$ with the number of unit cells $N$ and lattice vectors $\bR$.
The SOC-Hamiltonian is given by 
\begin{equation}
\begin{split}
H_{\textrm{SOC}}=\frac{1}{\hbar^2}\boldsymbol{\tilde{L}\cdot S}=\frac{1}{2\hbar}\boldsymbol{\tilde{L}\cdot \sigma}
\end{split}
\label{eq:H_SOC}
\end{equation}
with the Pauli matrices $\boldsymbol{\sigma}=(\sigma_x,\sigma_y,\sigma_z)$ and the modified angular momentum operator
\begin{equation}
\boldsymbol{\tilde{L}}=
\begin{pmatrix}
      \lambda_{\textrm{Mo}}\boldsymbol{L}_{l=2} & 0   &  0 \\
      0 & \lambda_{\textrm{S}}\boldsymbol{L}_{l=1}   & 0 \\
      0 & 0 &  \lambda_{\textrm{S}}\boldsymbol{L}_{l=1}
    \end{pmatrix}
\label{eq:L_op}
\end{equation}
that contains intra-atomic coupling parameters $\lambda_{\textrm{Mo}}$ for the $(l=2)$-subspace (d$_{z^2}$, d$_{xz}$, d$_{yz}$, d$_{x^2-y^2}$ and d$_{xy}$) and
$\lambda_{\textrm{S}} $ for the $(l=1)$-subspace (p$_x$, p$_y$ and p$_z$). In the given basis, the angular momentum algebra for spherical harmonics yields: 
\begin{equation}
\begin{split}
L_{x,l=1}&=
\begin{pmatrix}
      0 & 0   &  0 \\
      0 & 0   & -i\hbar \\
      0 & i\hbar & 0
    \end{pmatrix} \,,
L_{y,l=1}=
\begin{pmatrix}
      0 & 0   &  i\hbar \\
      0 & 0   & 0 \\
      -i\hbar & 0 & 0
    \end{pmatrix} \,, \\
L_{z,l=1}&= 
\begin{pmatrix}
      0 & -i\hbar   &  0 \\
      i\hbar & 0   & 0 \\
      0 & 0 & 0
    \end{pmatrix} 
\end{split}
\label{eq:L_1}
\end{equation}
and 
\begin{equation}
\begin{split}
L_{x,l=2}&=
\begin{pmatrix}
      0 & 0   &  \sqrt{3}i\hbar & 0 & 0 \\
      0 & 0   & 0 & 0 & i\hbar\\
      -\sqrt{3}i\hbar & 0   & 0 & -i\hbar & 0\\
      0 & 0   & i\hbar & 0 & 0\\
      0 & -i\hbar &  0 & 0 & 0
    \end{pmatrix}\,,
    \\
L_{y,l=2}&=
\begin{pmatrix}
      0 & -\sqrt{3}i\hbar   &  0 & 0 & 0 \\
      \sqrt{3}i\hbar & 0   & 0 & -i\hbar & 0\\
      0 & 0   & 0 & 0 & -i\hbar\\
      0 & i\hbar  & 0 & 0 & 0\\
      0 & 0 &  i\hbar & 0 & 0
    \end{pmatrix}\,,
    \\
L_{z,l=2}&= 
\begin{pmatrix}
      0 & 0   &  0 & 0 & 0 \\
      0 & 0   & -i\hbar & 0 & 0\\
      0 & i\hbar   & 0 & 0 & 0\\
      0 & 0   & 0 & 0 & -2i\hbar\\
      0 & 0 &  0 & 2i\hbar & 0
    \end{pmatrix}\,.
\end{split}
\label{eq:L_2}
\end{equation}
We choose the coupling constants $\lambda_{\textrm{Mo}}=90$ meV and $\lambda_{\textrm{S}}=20$ meV to reproduce the spin-orbit splittings at the K-point
as obtained from DFT calculations including spin-orbit coupling using fully relativistic pseudopotentials. The valence-band and conduction-band splittings
are $146$ meV and $3$ meV, respectively, with a like-spin ground state. Fig.~\ref{fig:DFT_bands} shows the excellent agreement between band structures directly from fully relativistic DFT calculations and from
diagonalization of the spin-augmented Wannier Hamiltonian (\ref{eq:H_tot}).
%
\\Starting from the density-density-like bare Coulomb interaction matrix elements in the Wannier basis,
%
\begin{equation}
\begin{split}
U_{\alpha\beta}(\bq)&=\sum_{\bR}e^{i\bq\cdot\bR}U_{\alpha\beta\beta\alpha}(\bR) \\
&=\sum_{\bR}e^{i\bq\cdot\bR}\bra{\boldsymbol{0},\alpha}\bra{\bR,\beta} U(\br,\br') \ket{\bR,\beta}\ket{\boldsymbol{0},\alpha}\,,
\end{split}
\label{eq:U_1}
\end{equation}
and the corresponding (statically) screened matrix elements $V_{\alpha\beta}(\bq)$, we obtain an analytic description of Coulomb interaction in freestanding monolayer TMDs that can be extended to include screening from a dielectric environment. To this end, we diagonalize the bare Coulomb matrix $\boldsymbol{U}(\bq)$ to obtain eigenvalues $U_i(\bq)$ and eigenvectors $\boldsymbol{e}_i(\bq)$. Since the momentum-dependence of the eigenvectors is weak, we use their long-wavelength limit in the following. The leading four eigenvalues are shown in Fig.~\ref{fig:U_fit}.
All further values are of similar size as $U_4(\bq)$. For the analytic description of the leading eigenvalue, we use
\begin{equation}
\begin{split}
U_1(q)=\frac{n_{\textrm{orb}} e^2}{2\varepsilon_0 A_{\textrm{UC}}}\frac{1}{q(1+\gamma q + \delta q^2 + \eta q^3)}\,,
\end{split}
\label{eq:U_1}
\end{equation}
where the area of the hexagonal unit cell $A_{\textrm{UC}}=\frac{\sqrt{3}}{2}a^2$ and the number of orbitals $n_{\textrm{orb}}=11$ ensure the proper normalization
of Coulomb matrix elements. 
The eigenvalues
$U_2(q)$ through $U_{11}(q)$ are fitted by third-order polynomials. The matrix elements of the screened interaction $\boldsymbol{V}(q)$ in the eigenbasis of the bare interaction are then obtained via
\begin{equation}
\begin{split}
V_i(q)=\varepsilon^{-1}_i(q)\,U_i(q)\,,
\end{split}
\label{eq:V_from_U}
\end{equation}
where the dielectric matrix $\boldsymbol{\varepsilon}(q)$ accounts for both the material-specific internal polarizability and the screening by the environment.
First, we introduce an analytic description for the freestanding monolayer dielectric function, i.e. in the absence of external screening.
While the eigenvalues $\varepsilon_2(q)$ through $\varepsilon_{11}(q)$ are again well described by third-order polynomials, the leading eigenvalue is expressed by
\begin{equation}
\begin{split}
\varepsilon_1(q)=\varepsilon_{\infty}(q)\frac{1-\beta^2 e^{-2qd}}{1+2\beta e^{-qd}+\beta^2 e^{-2qd}}\,,
\end{split}
\label{eq:eps_1}
\end{equation}
with $\beta= \frac{\varepsilon_{\infty}(q)-1}{\varepsilon_{\infty}(q)+1}$ \cite{rosner_wannier_2015} and the bulk dielectric constant given by a modified Resta model \cite{resta_thomas-fermi_1977}
\begin{equation}
\begin{split}
\varepsilon_{\infty}(q)=\frac{a+q^2}{\frac{a\, \textrm{sin}(qc)}{qbc}+q^2}+e \,.
\end{split}
\label{eq:eps_inf}
\end{equation}
As layer thickness, we use the layer separation in bulk $d=0.62$ nm \cite{kylanpaa_binding_2015}.
The leading four eigenvalues are shown in Fig.~\ref{fig:eps_fit}.
As soon as all fitting parameters are obtained, environmental screening can be taken into account according to the Wannier function continuum electrostatics approach \cite{rosner_wannier_2015} that combines a macroscopic electrostatic model for 
the screening by the dielectric environment with a localized description of Coulomb interaction.
The leading eigenvalue $\varepsilon_1(q)$, which is most sensitive to macroscopic screening, is modified by replacing the dielectric function of a freestanding monolayer 
with that of an arbitrary vertical heterostructure. The latter is obtained by solving Poisson's equation for a test charge in a slab with thickness $d$ and dielectric 
function $\varepsilon_{\infty}(q)$ embedded in a z-dependent dielectric profile \cite{florian_dielectric_2018}.
After calculating the eigenvalues $V_i(q)$, the Coulomb matrix in the Wannier basis is obtained by using the eigenvectors $\boldsymbol{e}_i(\bq)$. 
We finally compute screened Coulomb matrix elements 
in the Bloch-state representation by a unitary transformation using the coefficients $c^{\lambda}_{\alpha,\bk}$:
\begin{equation}
\begin{split}
V^{\lambda,\nu,\nu',\lambda'}_{\bk_1, \bk_2, \bk_3, \bk_4} = \frac{1}{N}\sum_{\alpha, \beta} \big( c_{\alpha, \bk_1}^{\lambda} \big)^{*} \big( c_{\beta, \bk_2}^{\nu} \big)^{*} V^{\alpha\beta}_{|\bk_1-\bk_4|}c_{\beta, \bk_3}^{\nu'} c_{\alpha, \bk_4}^{\lambda'}  \, ,
\end{split}
\label{eq:Coul_ME}
\end{equation}
where $\bk_4=\bk_1+\bk_2-\bk_3+\bG$ due to momentum conservation and the factor $1/N$ stems from the normalization of Wannier functions. Note that $V^{\alpha\beta}_{\bq}$ is periodic with respect to reciprocal lattice vectors $\bG$. In the following, we pull $1/(N A_{\textrm{UC}})=1/\mathcal{A}$ out of the matrix elements. 
\\For a TMD monolayer, the spin z component is a good quantum number due to the reflection symmetry by the x-y plane \cite{liu_three-band_2013, danovich_auger_2016}.
Every Bloch band can be assigned to $\ket{\uparrow}$ or $\ket{\downarrow}$. Since this symmetry can be slightly violated due to inaccuracies of our model, we enforce it by
assigning a definite spin to each band according to the dominant contribution given by the coefficients $c^{\lambda}_{\alpha,\bk}$. Furthermore, we make use of the fact that 
Coulomb interaction is spin-conserving, so that we can set Coulomb matrix elements $V^{\lambda,\nu,\nu',\lambda'}_{\bk_1 \bk_2 \bk_3 \bk_4}$ to zero if $\lambda$ and $\lambda'$ or $\nu$ and $\nu'$ belong to different spins.

\subsection{Derivation of microscopic EEA scattering rates}

We start from a Hamiltonian for Bloch electrons interacting via a statically screened Coulomb potential $V$:
\begin{equation}
 \begin{split}
H&=H_0+H_{\textrm{Coul}} \\&=
\sum_{\bk,c}\varepsilon^{c}_{\bk}a^{\dagger}_{\bk,c}a^{\phantom\dagger}_{\bk,c}+\sum_{\bk,v}\varepsilon^{v}_{\bk}a^{\dagger}_{\bk,v}a^{\phantom\dagger}_{\bk,v}\\
&+\frac{1}{2\mathcal{A}}\sum_{\substack{\bk,\bk',\bq \\ \lambda,\nu,\nu',\lambda'}}V^{\lambda,\nu,\nu',\lambda'}_{\bk,\bk',\bk'+\bq,\bk-\bq} 
a^{\dagger}_{\bk,\lambda}a^{\dagger}_{\bk',\nu}a^{\phantom\dagger}_{\bk'+\bq,\nu'}a^{\phantom\dagger}_{\bk-\bq,\lambda'}\,,
\end{split}
\label{eq:Hamiltonian}
\end{equation}
where $a^{\dagger}_{\bk,c/v}$ and $a^{\phantom\dagger}_{\bk,c/v}$ denote creation and annihilation operators, respectively, 
for a carrier with momentum $\bk$ in a conduction/valence band. 
The single-particle band structure is given by $\varepsilon^{\lambda}_{\bk}$.
To discuss quantities in an exciton picture, we introduce two-particle operators
\begin{equation}
\begin{split}
X^{\dagger}_{\alpha,\bq}&=\sum_{\bk,v,c}\phi^{vc}_{\alpha,\bk}(\bq)a_{\bk-\bq,\textrm{c}}^{\dagger} a_{\bk,\textrm{v}}^{\phantom\dagger}, \\
a_{\bk-\bq,\textrm{c}}^{\dagger} a_{\bk,\textrm{v}}^{\phantom\dagger}&=\sum_{\alpha}(\phi^{vc}_{\alpha,\bk}(\bq))^*X^{\dagger}_{\alpha,\bq}.
\end{split}
\label{eq:X_op}
\end{equation}
The wave functions $\phi^{vc}_{\alpha,\bk}(\bq)$ are solutions of the Bethe-Salpeter equation (BSE) in the absence of photoexcited carriers 
\begin{equation}
 \begin{split}
&(\varepsilon^{\textrm{c}}_{\bk-\bq}-\varepsilon^{\textrm{v}}_{\bk}-E_{\alpha,\bq})\phi^{vc}_{\alpha,\bk}(\bq) \\
-&\frac{1}{\mathcal{A}}\sum_{\bk',v',c'}
V^{c,v',v,c'}_{\bk-\bq,\bk',\bk,\bk'-\bq}\phi^{v'c'}_{\alpha,\bk'}(\bq)=0\,,
\end{split}
\label{eq:wannierSI}
\end{equation}
with two-particle eigenenergies $E_{\alpha,\bq}$ and the crystal area $\mathcal{A}$. The total momentum is denoted by $\bq$, 
while $\alpha$ is the quantum number that belongs to the relative motion of electron and hole. The wave functions fulfill
orthonormality and completeness relations:
\begin{equation}
\begin{split}
\sum_{\bk,v,c}(\wfc{v}{c}{\alpha}{\bk}{\bq})^*\wfc{v}{c}{\alpha'}{\bk}{\bq}&=\delta_{\alpha,\alpha'}, \\
\sum_{\alpha}\wfc{v}{c}{\alpha}{\bk}{\bq}(\wfc{v'}{c'}{\alpha}{\bk'}{\bq})^*&=\delta_{\bk,\bk'}\delta_{v,v'}\delta_{c,c'}\,.
\end{split}
\end{equation}
Auger-like EEA emerges as a higher-order carrier-carrier interaction process within the dynamics of microscopic exciton populations, 
which are described by two-particle correlation functions (doublets)
$n_{\alpha,\bq}= \Delta \big\langle X^{\dagger}_{\alpha,\bq}X^{\phantom\dagger}_{\alpha,\bq}  \big\rangle$ \cite{kira_many-body_2006}.
Correlation functions are defined recursively as the difference between an expectation value and all of its possible factorizations
into smaller correlation functions, i.e. with less operators \cite{kira_many-body_2006}. For example, a general doublet is given by
\begin{equation}
\begin{split}
\Delta &\big\langle a_{\bk-\bq,\textrm{1}}^{\dagger}a_{\bk'+\bq,\textrm{2}}^{\dagger}a_{\bk',\textrm{3}}^{\phantom\dagger} a_{\bk,\textrm{4}}^{\phantom\dagger} \big\rangle \\
=&\big\langle a_{\bk-\bq,\textrm{1}}^{\dagger}a_{\bk'+\bq,\textrm{2}}^{\dagger}a_{\bk',\textrm{3}}^{\phantom\dagger} a_{\bk,\textrm{4}}^{\phantom\dagger} \big\rangle \\
-&
\Delta \big\langle a_{\bk-\bq,\textrm{1}}^{\dagger}a_{\bk,\textrm{4}}^{\phantom\dagger}\big\rangle \Delta\big\langle a_{\bk'+\bq,\textrm{2}}^{\dagger}a_{\bk',\textrm{3}}^{\phantom\dagger}\big\rangle\delta_{\bq,\boldsymbol{0}} \\
+&\Delta \big\langle a_{\bk-\bq,\textrm{1}}^{\dagger}a_{\bk',\textrm{3}}^{\phantom\dagger}\big\rangle \Delta\big\langle a_{\bk'+\bq,\textrm{2}}^{\dagger}a_{\bk,\textrm{4}}^{\phantom\dagger}\big\rangle
\delta_{\bq,\bk-\bk'}\,.
\end{split}
\label{eq:factorize}
\end{equation}
Note that translational invariance of the crystal enforces momentum conservation within any expectation value and that transposition of fermionic operators
yields a minus sign. The dynamics of operators is
governed by the Heisenberg EOM:
\begin{equation}
 \begin{split}
i\hbar\frac{d}{dt} \hat{A}= \left[ \hat{A}, H \right]\,.
\end{split}
\label{eq:heisenberg}
\end{equation}
We apply the so-called cluster expansion technique \cite{kira_many-body_2006} to formulate the dynamical equations in terms of correlation functions instead of expectation values.
To this end, we make use of the identity
\begin{equation}
 \begin{split}
\left[ \hat{A}\hat{B}, \hat{C} \right]=\hat{A}\left[ \hat{B}, \hat{C} \right]_{+}-\left[ \hat{A}, \hat{C} \right]_+\hat{B}\,,
\end{split}
\label{eq:heisenberg}
\end{equation}
where $\left[ \cdot\,,\cdot \right]_{+} $ denotes the anti-commutator.
\\ The $n_{\alpha,\bq}$ are transformed to the Bloch representation using the expansion of two-particle operators (\ref{eq:X_op}):
\begin{equation}
\begin{split}
&n_{\alpha,\bq}= \Delta\left<X^{\dagger}_{\alpha,\bq} X^{\phantom\dagger}_{\alpha,\bq}\right>
 \\ &= \sum_{\substack{\bk,v,c \\ \bk',v',c'}}\wfc{v'}{c'}{\alpha}{\bk'}{\bq}(\wfc{v}{c}{\alpha}{\bk}{\bq})^*
\Delta\left<a^{\dagger}_{\bk'-\bq,c'}a^{\phantom\dagger}_{\bk',v'}
a^{\dagger}_{\bk,v}a^{\phantom\dagger}_{\bk-\bq,c}\right> \\
 & = \sum_{\substack{\bk,v,c \\ \bk',v',c'}}\wfc{v'}{c'}{\alpha}{\bk'}{\bq}(\wfc{v}{c}{\alpha}{\bk}{\bq})^*
\Delta\left<a^{\dagger}_{\bk,v}a^{\dagger}_{\bk'-\bq,c'}a^{\phantom\dagger}_{\bk',v'}
a^{\phantom\dagger}_{\bk-\bq,c}\right> \\
 & = \sum_{\substack{\bk,v,c \\ \bk',v',c'}}\wfc{v'}{c'}{\alpha}{\bk'+\bq}{\bq}(\wfc{v}{c}{\alpha}{\bk}{\bq})^*
\Delta\left<a^{\dagger}_{\bk,v}a^{\dagger}_{\bk',c'}a^{\phantom\dagger}_{\bk'+\bq,v'}
a^{\phantom\dagger}_{\bk-\bq,c}\right>\,.
\end{split}
\end{equation}
Within a correlation function, transposition of two fermionic operators yields a minus sign \cite{kira_many-body_2006}.
The time derivative of the carrier-carrier correlation in the Bloch picture can be evaluated by commutating with the Hamiltonian (\ref{eq:Hamiltonian}).
Since we are interested in the dynamics due to EEA, we retain only coupling terms to three-particle correlation functions (triplets). The other terms
are discussed at length in Ref.~\cite{kira_many-body_2006}.
This yields:
\begin{equation}
\begin{split}
i\hbar&\frac{d}{d t}n_{\alpha,\bq}
 = \sum_{\bk',v',c'}\wfc{v'}{c'}{\alpha}{\bk'+\bq}{\bq}\sum_{\bk,v,c}(\wfc{v}{c}{\alpha}{\bk}{\bq})^* \frac{1}{\mathcal{A}}\sum_{\bl,\bl',1,2,3}\times \\ 
 \Bigg\{&V^{c,1,2,3}_{\bk-\bq,\bl,\bl+\bl',\bk-\bq-\bl'} 
 \Delta\left<a^{\dagger}_{\bk,v}a^{\dagger}_{\bk',c'}a^{\dagger}_{\bl,1}a^{\phantom\dagger}_{\bl+\bl',2}a^{\phantom\dagger}_{\bk'+\bq,v'}a^{\phantom\dagger}_{\bk-\bq-\bl',3}           \right> \\
 +&V^{v',1,2,3}_{\bk'+\bq,\bl,\bl+\bl',\bk'+\bq-\bl'} 
 \Delta\left<a^{\dagger}_{\bk,v}a^{\dagger}_{\bk',c'}a^{\dagger}_{\bl,1}a^{\phantom\dagger}_{\bl+\bl',2}a^{\phantom\dagger}_{\bk'+\bq-\bl',3}a^{\phantom\dagger}_{\bk-\bq,c}           \right> \\
 -(&V^{c',1,2,3}_{\bk',\bl,\bl+\bl',\bk'-\bl'} 
 \Delta\left<a^{\dagger}_{\bk-\bq,c}a^{\dagger}_{\bk'+\bq,v'}a^{\dagger}_{\bl,1}a^{\phantom\dagger}_{\bl+\bl',2}a^{\phantom\dagger}_{\bk'-\bl',3}a^{\phantom\dagger}_{\bk,v}           \right>)^* \\
 -(&V^{v,1,2,3}_{\bk,\bl,\bl+\bl',\bk-\bl'} 
 \Delta\left<a^{\dagger}_{\bk-\bq,c}a^{\dagger}_{\bk'+\bq,v'}a^{\dagger}_{\bl,1}a^{\phantom\dagger}_{\bl+\bl',2}a^{\phantom\dagger}_{\bk',c'}a^{\phantom\dagger}_{\bk-\bl',3}           \right>)^* \Bigg\}\,.
\end{split}
\label{eq:n_to_triplet}
\end{equation}
We keep all band indices $1,2,3$ that allow for a pairing of conduction and valence band operators into two-particle operators, 
which yields three possible combinations for each of the terms in Eq.~(\ref{eq:n_to_triplet}). After re-introducing two-particle
operators and using the orthonormality relations for wave functions, the 12 resulting terms can by rearranged as follows:
\begin{equation}
\begin{split}
& i\hbar\frac{d}{dt}n_{\alpha,\bq}= 
 \frac{1}{\mathcal{A}}\sum_{\bl'}
 \times \\
 & 2i\,\textrm{Im}\Bigg\{\sum_{\beta\delta}\Big(\hat{V}^{(1),D}_{\alpha,\beta,\delta,\bq     , \bl'}\Big)^*\Delta\left<X^{\dagger}_{\alpha,\bq}X^{\phantom\dagger}_{\beta,\bq+\bl'}X^{\phantom\dagger}_{\delta,-\bl'} \right> \\
 +&\sum_{\beta\delta}    (\hat{V}^{(1),D}_{\beta,\alpha,\delta,\bq+\bl',-\bl'} - \hat{V}^{(1),X}_{\beta,\alpha,\delta,\bq+\bl',-\bl'} )      \Big(\Delta\left<X^{\dagger}_{\beta,\bq+\bl'}X^{\phantom\dagger}_{\alpha,\bq}X^{\phantom\dagger}_{\delta,\bl'}  \right>\Big)^* \\ 
 +&\sum_{\beta\delta}\Big((\hat{V}^{(2),D}_{\beta,\alpha,\delta,\bq     , \bl'}-\hat{V}^{(2),X}_{\beta,\alpha,\delta,\bq     , \bl'})\Big)^* \Delta\left<X^{\dagger}_{\beta,\bq-\bl'}X^{\phantom\dagger}_{\alpha,\bq}X^{\phantom\dagger}_{\delta,-\bl'}       \right> \\
 +&\sum_{\beta\delta} \hat{V}^{(2),D}_{\alpha,\beta,\delta,\bq-\bl'     , -\bl'}    \Big( \Delta\left<X^{\dagger}_{\alpha,\bq}X^{\phantom\dagger}_{\beta,\bq-\bl'}X^{\phantom\dagger}_{\delta,\bl'}         \right>\Big)^*
 \Bigg\}\,,
\end{split}
\label{eq:EOM_n_X}
\end{equation}
where we have introduced effective direct (D) and exchange (X) exciton-exciton interaction matrix elements 
$V^{D/X,\alpha,\beta,\delta}_{\bq,-\bl}=V^{(1),D/X}_{\alpha,\beta,\delta,\bq,-\bl} - 
 V^{(2),D/X}_{\alpha,\beta,\delta,\bq-\bl,-\bl}$ as defined in Eq.~\eqref{eq:coul_me}, where $V^{(1)}$ and $V^{(2)}$ describe Auger-like scattering of
electrons and holes, respectively.

%

\begin{figure}
\centering
\includegraphics[width=\columnwidth]{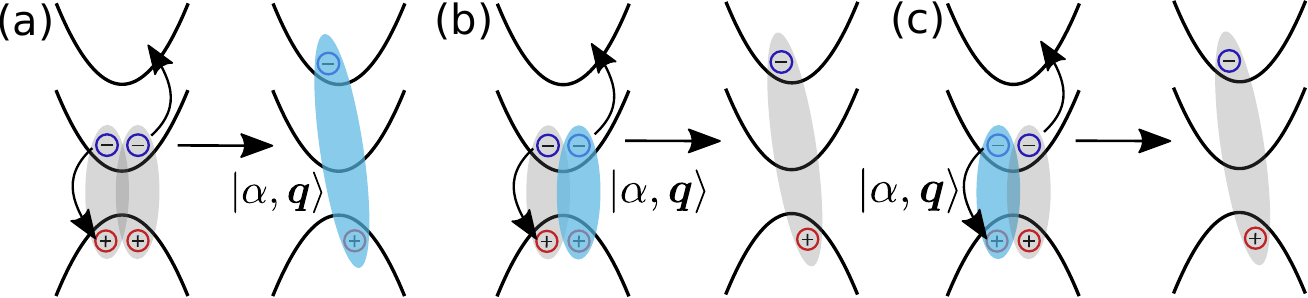}
\caption{Different roles of exciton state $\big|\alpha,\boldsymbol{q}\big>$ in EEA process. (a): As high-energy target state. (b): As 1s-exciton state that takes up the excess energy and momentum from the annihilated exciton. (c) As the annihilated 1s-exciton.}
\label{fig:three_channels}
\end{figure}

%
Similar as for exciton populations, we use the Heisenberg equation to derive EOM for triplets:
\begin{equation}
\begin{split}
&i\hbar\frac{d}{dt}\Delta\left<X^{\dagger}_{\alpha,\bq_1}X^{\phantom\dagger}_{\beta,\bq_2}X^{\phantom\dagger}_{\delta,-\bl'}  \right> \\
= & \sum_{\bk,\bk',\bl}\sum_{v,c,v',c',v'',c''}
\wfc{v}{c}{\alpha}{\bk}{\bq_1}(\wfc{v'}{c'}{\beta}{\bk'}{\bq_2})^*(\wfc{v''}{c''}{\delta}{\bl}{-\bl'})^*\times\\
&i\hbar\frac{d}{dt}
\Delta\left<  a^{\dagger}_{\bk-\bq_1,c}a^{\dagger}_{\bk',v'}a^{\dagger}_{\bl,v''}a^{\phantom\dagger}_{\bl+\bl',c''}a^{\phantom\dagger}_{\bk'-\bq_2,c'}a^{\phantom\dagger}_{\bk,v}     \right>
\end{split}
\label{eq:triplet_EOM}
\end{equation}
with
\begin{widetext}
\begin{equation}
\begin{split}
&i\hbar\frac{d}{dt}
\Delta\left<  a^{\dagger}_{\bk-\bq_1,c}a^{\dagger}_{\bk',v'}a^{\dagger}_{\bl,v''}a^{\phantom\dagger}_{\bl+\bl',c''}a^{\phantom\dagger}_{\bk'-\bq_2,c'}a^{\phantom\dagger}_{\bk,v}     \right>\\
&=\big(\en{v}{\bk}+\en{c'}{\bk'-\bq_2}+\en{c''}{\bl+\bl'}-\en{v''}{\bl}-\en{v'}{\bk'}-\en{c}{\bk-\bq_1}  \big)
\Delta\left<  a^{\dagger}_{\bk-\bq_1,c}a^{\dagger}_{\bk',v'}a^{\dagger}_{\bl,v''}a^{\phantom\dagger}_{\bl+\bl',c''}a^{\phantom\dagger}_{\bk'-\bq_2,c'}a^{\phantom\dagger}_{\bk,v}     \right>\\
&- \frac{1}{\mathcal{A}} \sum_{\bp,\bp',1,2,3} 
V^{1,2,3,c}_{\bp,\bp',\bp+\bp'-(\bk-\bq_1),\bk-\bq_1}\left< a^{\dagger}_{\bp,1}a^{\dagger}_{\bp',2} a^{\phantom\dagger}_{\bp+\bp'-(\bk-\bq_1),3}a^{\dagger}_{\bk',v'}a^{\dagger}_{\bl,v''}a^{\phantom\dagger}_{\bl+\bl',c''}a^{\phantom\dagger}_{\bk'-\bq_2,c'}a^{\phantom\dagger}_{\bk,v}     \right>  \\
&- \frac{1}{\mathcal{A}} \sum_{\bp,\bp',1,2,3} 
V^{1,2,3,v'}_{\bp,\bp',\bp+\bp'-\bk',\bk'}\left<a^{\dagger}_{\bk-\bq_1,c} a^{\dagger}_{\bp,1}a^{\dagger}_{\bp',2} a^{\phantom\dagger}_{\bp+\bp'-\bk',3}a^{\dagger}_{\bl,v''}a^{\phantom\dagger}_{\bl+\bl',c''}a^{\phantom\dagger}_{\bk'-\bq_2,c'}a^{\phantom\dagger}_{\bk,v}     \right>  \\
&- \frac{1}{\mathcal{A}} \sum_{\bp,\bp',1,2,3} 
V^{1,2,3,v''}_{\bp,\bp',\bp+\bp'-\bl,\bl}\left<a^{\dagger}_{\bk-\bq_1,c} a^{\dagger}_{\bk',v'}a^{\dagger}_{\bp,1}a^{\dagger}_{\bp',2} a^{\phantom\dagger}_{\bp+\bp'-\bl,3}a^{\phantom\dagger}_{\bl+\bl',c''}a^{\phantom\dagger}_{\bk'-\bq_2,c'}a^{\phantom\dagger}_{\bk,v}     \right>  \\
&+ \frac{1}{\mathcal{A}} \sum_{\bp,\bp',1,2,3} 
V^{c'',1,2,3}_{\bl+\bl',\bp,\bp+\bp',\bl+\bl'-\bp'}\left<a^{\dagger}_{\bk-\bq_1,c}a^{\dagger}_{\bk',v'}a^{\dagger}_{\bl,v''}a^{\dagger}_{\bp,1}a^{\phantom\dagger}_{\bp+\bp',2} a^{\phantom\dagger}_{\bl+\bl'-\bp',3}a^{\phantom\dagger}_{\bk'-\bq_2,c'}a^{\phantom\dagger}_{\bk,v}     \right>  \\
&+ \frac{1}{\mathcal{A}} \sum_{\bp,\bp',1,2,3} 
V^{c',1,2,3}_{\bk'-\bq_2,\bp,\bp+\bp',\bk'-\bq_2-\bp'}
\left<a^{\dagger}_{\bk-\bq_1,c}a^{\dagger}_{\bk',v'}a^{\dagger}_{\bl,v''}a^{\phantom\dagger}_{\bl+\bl',c''}a^{\dagger}_{\bp,1}a^{\phantom\dagger}_{\bp+\bp',2} a^{\phantom\dagger}_{\bk'-\bq_2-\bp',3}a^{\phantom\dagger}_{\bk,v}     \right>  \\
&+ \frac{1}{\mathcal{A}} \sum_{\bp,\bp',1,2,3} 
V^{v,1,2,3}_{\bk,\bp,\bp+\bp',\bk-\bp'}
\left<a^{\dagger}_{\bk-\bq_1,c}a^{\dagger}_{\bk',v'}a^{\dagger}_{\bl,v''}a^{\phantom\dagger}_{\bl+\bl',c''}a^{\phantom\dagger}_{\bk'-\bq_2,c'}a^{\dagger}_{\bp,1}a^{\phantom\dagger}_{\bp+\bp',2} a^{\phantom\dagger}_{\bk-\bp',3} \right>  \\
& - i\hbar\frac{d}{dt}
\Delta\left<  a^{\dagger}_{\bk-\bq_1,c}a^{\dagger}_{\bk',v'}a^{\dagger}_{\bl,v''}a^{\phantom\dagger}_{\bl+\bl',c''}a^{\phantom\dagger}_{\bk'-\bq_2,c'}a^{\phantom\dagger}_{\bk,v}     \right>\Big|_{\textrm{factorizations}}\,.
\end{split}
\end{equation}
\end{widetext}
Here, the first term leading to oscillations with free energies stems from commutating with the Hamiltonian $H_0$, while the coupling to four-particle expectation values
is due to the Coulomb interaction Hamiltonian $H_{\textrm{Coul}}$. 
Since we consider the time derivative of a correlation function, we have to subtract all terms that are due to factorizations into smaller correlation functions similar to
Eq.~(\ref{eq:factorize}).
The four-particle expectation values themselves can be represented by correlation functions according to the scheme
\begin{equation}
\begin{split}
\left<4\right>&=\left<1\right>\left<1\right>\left<1\right>\left<1\right>+
\left<1\right>\left<1\right>\Delta\left<2\right>+\Delta\left<2\right>\Delta\left<2\right> \\
&+\left<1\right>\Delta\left<3\right>+\Delta\left<4\right>\,.
\end{split}
\end{equation}
We discard four-particle correlations to truncate the hierarchy and obtain a closed set of equations. Moreover, we discard 
one-particle quantities $\Delta\big\langle a^{\dagger}_{\bk,\lambda}a^{\phantom\dagger}_{\bk,\lambda'}  \big\rangle$ that are not band-diagonal 
as well as conduction-band populations $f^{c}_{\bk}=\Delta\big\langle a^{\dagger}_{\bk,c}a^{\phantom\dagger}_{\bk,c}  \big\rangle$, while we approximate
valence-band populations by $1$. By neglecting electron and hole populations, we assume that corrections due to the occupation of single-particle phase space by excitons is small. 
Among the doublets, we retain those which correspond to exciton populations of the states $\ket{\alpha,\bq_1}$, $\ket{\beta,\bq_2}$ or $\ket{\delta,-\bl'}$. 
Only for these factorizations, it will be possible to introduce effective interaction matrix elements as in Eq.~(\ref{eq:EOM_n_X}). We therefore assume that all further factorizations
are beyond exciton-exciton scattering, e.g. by mixing exciton states.
After cancellation
between factorizations of the four-particle expectation values and the time derivative of the factorizations of triplets, we obtain:
\begin{widetext}
\begin{equation}
\begin{split}
&i\hbar\frac{d}{dt}
\Delta\left<  a^{\dagger}_{\bk-\bq_1,c}a^{\dagger}_{\bk',v'}a^{\dagger}_{\bl,v''}a^{\phantom\dagger}_{\bl+\bl',c''}a^{\phantom\dagger}_{\bk'-\bq_2,c'}a^{\phantom\dagger}_{\bk,v}     \right> \\
=
& \big(\en{v}{\bk}+\en{c'}{\bk'-\bq_2}+\en{c''}{\bl+\bl'}-\en{v''}{\bl}-\en{v'}{\bk'}-\en{c}{\bk-\bq_1}  \big)\times \\
&\Delta\left<  a^{\dagger}_{\bk-\bq_1,c}a^{\dagger}_{\bk',v'}a^{\dagger}_{\bl,v''}a^{\phantom\dagger}_{\bl+\bl',c''}a^{\phantom\dagger}_{\bk'-\bq_2,c'}a^{\phantom\dagger}_{\bk,v}  \right>  \\
 & + \frac{1}{\mathcal{A}}\sum_{\bl'',v''',c'''} V^{v,c''',c,v'''}_{\bk,\bl''-\bq_1,\bk-\bq_1,\bl''}
 \Delta\left<  a^{\dagger}_{\bl''-\bq_1,c'''}a^{\dagger}_{\bk',v'}a^{\dagger}_{\bl,v''}a^{\phantom\dagger}_{\bl+\bl',c''}a^{\phantom\dagger}_{\bk'-\bq_2,c'}a^{\phantom\dagger}_{\bl'',v'''}     \right>\\
 & - \frac{1}{\mathcal{A}}\sum_{\bl'',v''',c'''} V^{c',v''',v',c'''}_{\bk'-\bq_2,\bl'',\bk',\bl''-\bq_2}
 \Delta\left<  a^{\dagger}_{\bk-\bq_1,c}a^{\dagger}_{\bl'',v'''}a^{\dagger}_{\bl,v''}a^{\phantom\dagger}_{\bl+\bl',c''}a^{\phantom\dagger}_{\bl''-\bq_2,c'''}a^{\phantom\dagger}_{\bk,v}     \right>\\
 & - \frac{1}{\mathcal{A}}\sum_{\bl'',v''',c'''} V^{c'',v''',v'',c'''}_{\bl+\bl',\bl'',\bl,\bl''+\bl'}
 \Delta\left<  a^{\dagger}_{\bk-\bq_1,c}a^{\dagger}_{\bk',v'}a^{\dagger}_{\bl'',v'''}a^{\phantom\dagger}_{\bl''+\bl',c'''}a^{\phantom\dagger}_{\bk'-\bq_2,c'}a^{\phantom\dagger}_{\bk,v}        \right> \\
  &+\frac{1}{\mathcal{A}} \sum_{\bp,1,2,3}\Bigg\{ \\
 &\Big[V^{1,2,3,c}_{\bp,\bk-\bq_1-\bl',\bp-\bl',\bk-\bq_1} -  V^{2,1,3,c}_{\bk-\bq_1-\bl',\bp,\bp-\bl',\bk-\bq_1} \Big]
 \Delta\left<  a^{\dagger}_{\bk-\bq_1-\bl',2}a^{\dagger}_{\bk',v'}a^{\phantom\dagger}_{\bk'-\bq_2,c'}a^{\phantom\dagger}_{\bk,v}     \right> 
 \Delta\left<  a^{\dagger}_{\bp,1}a^{\dagger}_{\bl,v''}a^{\phantom\dagger}_{\bl+\bl',c''}a^{\phantom\dagger}_{\bp-\bl',3}    \right>\\
  +&\Big[V^{1,2,3,c}_{\bp,\bk+\bl',\bp+\bl'+\bq_1,\bk-\bq_1} -  V^{2,1,3,c}_{\bk+\bl',\bp,\bp+\bl'+\bq_1,\bk-\bq_1} \Big]
 \Delta\left<  a^{\dagger}_{\bk+\bl',2}a^{\dagger}_{\bl,v''}a^{\phantom\dagger}_{\bl+\bl',c''}a^{\phantom\dagger}_{\bk,v}     \right> 
 \Delta\left<  a^{\dagger}_{\bp,1}a^{\dagger}_{\bk',v'}a^{\phantom\dagger}_{\bk'-\bq_2,c'}a^{\phantom\dagger}_{\bp+\bq_2,3}    \right>\\
 -&\Big[V^{1,2,3,v'}_{\bp,\bk'-\bl',\bp-\bl',\bk'} -  V^{2,1,3,v'}_{\bk'-\bl',\bp,\bp-\bl',\bk'} \Big]
 \Delta\left<  a^{\dagger}_{\bk'-\bl',2}a^{\dagger}_{\bk-\bq_1,c}a^{\phantom\dagger}_{\bk'-\bq_2,c'}a^{\phantom\dagger}_{\bk,v}     \right> 
 \Delta\left<  a^{\dagger}_{\bp,1}a^{\dagger}_{\bl,v''}a^{\phantom\dagger}_{\bl+\bl',c''}a^{\phantom\dagger}_{\bp-\bl',3}    \right> \\
+&\Big[V^{c',1,2,3}_{\bk'-\bq_2,\bp,\bk'-\bq_2+\bl',\bp-\bl'} -  V^{c',1,3,2}_{\bk'-\bq_2,\bp,\bp-\bl',\bk'-\bq_2+\bl'} \Big]
 \Delta\left<  a^{\dagger}_{\bk-\bq_1,c}a^{\dagger}_{\bk',v'}a^{\phantom\dagger}_{\bk,v}a^{\phantom\dagger}_{\bk'-\bq_2+\bl',2}     \right> 
 \Delta\left<  a^{\dagger}_{\bp,1}a^{\dagger}_{\bl,v''}a^{\phantom\dagger}_{\bl+\bl',c''}a^{\phantom\dagger}_{\bp-\bl',3}    \right>\\
-&\Big[V^{v,1,2,3}_{\bk,\bp,\bk-\bq_1-\bl',\bp+\bq_2} -  V^{v,1,3,2}_{\bk,\bp,\bp+\bq_2,\bk-\bq_1-\bl'} \Big]
 \Delta\left<  a^{\dagger}_{\bk-\bq_1,c}a^{\dagger}_{\bl,v''}a^{\phantom\dagger}_{\bl+\bl',c''}a^{\phantom\dagger}_{\bk-\bq_1-\bl',2}     \right> 
 \Delta\left<  a^{\dagger}_{\bp,1}a^{\dagger}_{\bk',v'}a^{\phantom\dagger}_{\bk'-\bq_2,c'}a^{\phantom\dagger}_{\bp+\bq_2,3}    \right>\\
  -&\Big[V^{v,1,2,3}_{\bk,\bp,\bk+\bl',\bp-\bl'} -  V^{v,1,3,2}_{\bk,\bp,\bp-\bl',\bk+\bl'} \Big]
 \Delta\left<  a^{\dagger}_{\bk-\bq_1,c}a^{\dagger}_{\bk',v'}a^{\phantom\dagger}_{\bk'-\bq_2,c'}a^{\phantom\dagger}_{\bk+\bl',2}     \right> 
 \Delta\left<  a^{\dagger}_{\bp,1}a^{\dagger}_{\bl,v''}a^{\phantom\dagger}_{\bl+\bl',c''}a^{\phantom\dagger}_{\bp-\bl',3}    \right> \\
  +&\Big[V^{1,2,3,v''}_{\bp,\bl+\bl'+\bq_1,\bp+\bq_2,\bl} -  V^{2,1,3,v''}_{\bl+\bl'+\bq_1,\bp,\bp+\bq_2,\bl} \Big]
 \Delta\left<  a^{\dagger}_{\bl+\bl'+\bq_1,2}a^{\dagger}_{\bk-\bq_1,c}a^{\phantom\dagger}_{\bk,v}a^{\phantom\dagger}_{\bl+\bl',c''}     \right> 
 \Delta\left<  a^{\dagger}_{\bp,1}a^{\dagger}_{\bk',v'}a^{\phantom\dagger}_{\bk'-\bq_2,c'}a^{\phantom\dagger}_{\bp+\bq_2,3}    \right> \\
  -&\Big[V^{c'',1,2,3}_{\bl+\bl',\bp,\bl-\bq_1,\bp+\bq_2} -  V^{c'',1,3,2}_{\bl+\bl',\bp,\bp+\bq_2,\bl-\bq_1} \Big]
 \Delta\left<  a^{\dagger}_{\bk-\bq_1,c}a^{\dagger}_{\bl,v''}a^{\phantom\dagger}_{\bl-\bq_1,2}a^{\phantom\dagger}_{\bk,v}     \right> 
 \Delta\left<  a^{\dagger}_{\bk',v'}a^{\dagger}_{\bp,1}a^{\phantom\dagger}_{\bp+\bq_2,3}a^{\phantom\dagger}_{\bk'-\bq_2,c'}    \right>\Bigg\} \\
&+\sum_{1}\Big[V^{c'',1,v'',v'}_{\bl+\bl',\bk'-\bl',\bl,\bk'} -  V^{1,c'',v'',v'}_{\bk'-\bl',\bl+\bl',\bl,\bk'} \Big]
 \Delta\left<  a^{\dagger}_{\bk'-\bl',1}a^{\dagger}_{\bk-\bq_1,c}a^{\phantom\dagger}_{\bk,v}a^{\phantom\dagger}_{\bk'-\bq_2,c'}     \right> \\
&+\sum_{1}\Big[V^{c'',c',1,v'}_{\bl+\bl',\bk'-\bq_2,\bl+\bl'-\bq_2,\bk'} -  V^{c',c'',1,v'}_{\bk'-\bq_2,\bl+\bl',\bl+\bl'-\bq_2,\bk'} \Big]
 \Delta\left<  a^{\dagger}_{\bk-\bq_1,c}a^{\dagger}_{\bl,v''}a^{\phantom\dagger}_{\bk,v}a^{\phantom\dagger}_{\bl+\bl'-\bq_2,1}     \right>  \\
&+\sum_{1}\Big[V^{c',1,v'',v'}_{\bk'-\bq_2,\bl+\bq_2,\bl,\bk'} - V^{1,c',v'',v'}_{\bl+\bq_2,\bk'-\bq_2,\bl,\bk'} \Big]
 \Delta\left<  a^{\dagger}_{\bl+\bq_2,1}a^{\dagger}_{\bk-\bq_1,c}a^{\phantom\dagger}_{\bl+\bl',c''}a^{\phantom\dagger}_{\bk,v}     \right> \\
&+\sum_{1}\Big[V^{c'',c',1,v''}_{\bl+\bl',\bk'-\bq_2,\bk'+\bl'-\bq_2,\bl} - V^{c',c'',1,v''}_{\bk'-\bq_2,\bl+\bl',\bk'+\bl'-\bq_2,\bl} \Big]
 \Delta\left<  a^{\dagger}_{\bk',v'}a^{\dagger}_{\bk-\bq_1,c}a^{\phantom\dagger}_{\bk,v}a^{\phantom\dagger}_{\bk'+\bl'-\bq_2,1}     \right> \,.
\end{split}
\label{eq:three_particle_bloch}
\end{equation}
\end{widetext}
The above result is inserted into Eq.~(\ref{eq:triplet_EOM}). After re-introducing two-particle operators, using the orthonormality of wave functions as 
well as the Bethe-Salpeter equation~(\ref{eq:wannierSI}) and discarding all doublets that are not density-like, we arrive at:
\begin{widetext}
\begin{equation}
\begin{split}
&i\hbar\frac{d}{dt}\Delta\left<X^{\dagger}_{\alpha,\bq_1}X^{\phantom\dagger}_{\beta,\bq_2}X^{\phantom\dagger}_{\delta,-\bl'}  \right> \\
= & \big(E_{\beta,\bq_2} + E_{\delta,-\bl'} - E_{\alpha,\bq_1} \big)\Delta\left<X^{\dagger}_{\alpha,\bq_1}X^{\phantom\dagger}_{\beta,\bq_2}X^{\phantom\dagger}_{\delta,-\bl'}  \right>\\
+ \frac{1}{\mathcal{A}}\Big\{
&\Big[\hat{V}^{(1),X}_{\alpha,\beta,\delta,\bq_1,\bl'} - \hat{V}^{(1),D}_{\alpha,\beta,\delta,\bq_1,\bl'} \Big]
 n_{\beta,\bq_2} n_{\delta,-\bl'}
 +\Big[\hat{V}^{(2),X}_{\alpha,\beta,\delta,\bq_2,\bl'} - \hat{V}^{(2),D}_{\alpha,\beta,\delta,\bq_2,\bl'} \Big]
n_{\alpha,\bq_1} n_{\delta,-\bl'} \\
+ &\Big[\hat{V}^{(1),X}_{\alpha,\delta,\beta,\bq_1,-\bq_2} - \hat{V}^{(1),D}_{\alpha,\delta,\beta,\bq_1,-\bq_2} \Big]
 n_{\beta,\bq_2} n_{\delta,-\bl'}
 - \Big[\hat{V}^{(1),X}_{\alpha,\beta,\delta,\bq_1,\bl'} - \hat{V}^{(1),D}_{\alpha,\beta,\delta,\bq_1,\bl'} \Big]
 n_{\alpha,\bq_1} n_{\delta,-\bl'} \\ -& 
 \Big[\hat{V}^{(2),X}_{\alpha,\beta,\delta,\bq_2,\bl'} - \hat{V}^{(2),D}_{\alpha,\beta,\delta,\bq_2,\bl'} \Big]
 n_{\beta,\bq_2}    n_{\delta,-\bl'}
  - \Big[\hat{V}^{(2),X}_{\alpha,\delta,\beta,-\bl',-\bq_2} - \hat{V}^{(2),D}_{\alpha,\delta,\beta,-\bl',-\bq_2} \Big]
 n_{\beta,\bq_2}    n_{\delta,-\bl'} \\
 +& \Big[\hat{V}^{(2),X}_{\alpha,\delta,\beta,-\bl',-\bq_2} - \hat{V}^{(2),D}_{\alpha,\delta,\beta,-\bl',-\bq_2} \Big]   n_{\alpha,\bq_1}    n_{\beta,\bq_2}  
  - \Big[\hat{V}^{(1),X}_{\alpha,\delta,\beta,\bq_1,-\bq_2} - \hat{V}^{(1),D}_{\alpha,\delta,\beta,\bq_1,-\bq_2} \Big]  n_{\alpha,\bq_1}    n_{\beta,\bq_2} \\
+&\Big[\hat{V}^{(2),X}_{\alpha,\beta,\delta,\bq_2,\bl'} - \hat{V}^{(2),D}_{\alpha,\beta,\delta,\bq_2,\bl'} \Big]
n_{\alpha,\bq_1}
- \Big[\hat{V}^{(1),X}_{\alpha,\delta,\beta,\bq_1,-\bq_2} - \hat{V}^{(1),D}_{\alpha,\delta,\beta,\bq_1,-\bq_2} \Big]  n_{\alpha,\bq_1}    \\
+& \Big[\hat{V}^{(2),X}_{\alpha,\delta,\beta,-\bl',-\bq_2} - \hat{V}^{(2),D}_{\alpha,\delta,\beta,-\bl',-\bq_2} \Big]   n_{\alpha,\bq_1}
- \Big[\hat{V}^{(1),X}_{\alpha,\beta,\delta,\bq_1,\bl'} - \hat{V}^{(1),D}_{\alpha,\beta,\delta,\bq_1,\bl'} \Big]
 n_{\alpha,\bq_1}
 \Big\}\,.
\end{split}
\label{eq:triplet_final}
\end{equation}
\end{widetext}
Note that by focussing on terms that contain exciton populations, the Coulomb matrix elements are automatically forced to Auger-like band index combinations
containing one inter-gap scattering process. This means that on the triplet level, the only Coulomb interaction
processes between exciton densities are of EEA type. Any scattering processes that conserve the number of excitons would involve terms that are
of third order in the exciton density, which require the factorization of six-particle expectation values and therefore appear on a higher
level of cluster expansion.
\\We eliminate the triplets by adding a phenomenological damping $\Gamma$ to the oscillating terms and applying
the Markov approximation \cite{kira_many-body_2006}, which yields algebraic expressions:
\begin{equation}
\begin{split}
\Delta\left<X^{\dagger}_{\alpha,\bq_1}X^{\phantom\dagger}_{\beta,\bq_2}X^{\phantom\dagger}_{\delta,-\bl'}  \right>(t)
\approx  \frac{-K_{\alpha,\beta,\delta,\bq_1,\bq_2,-\bl'}(t)}{E_{\beta,\bq_2} + E_{\delta,-\bl'} - E_{\alpha,\bq_1}-i\Gamma }
 \,,
\end{split}
\end{equation}
where $K(t)$ denotes the inhomogeneity on the RHS of Eq.~(\ref{eq:triplet_final}). The triplets in Markov approximation
are inserted into Eq.~(\ref{eq:EOM_n_X}). After combining the terms and using the relation $\textrm{Im}\,z^*=-\textrm{Im}\,z$, we arrive at
the final result:
\begin{equation}
 \begin{split}
&\frac{d}{dt} n_{\alpha,\bq}= 
 \frac{1}{\mathcal{A}^2}\sum_{\bl}
 \frac{2}{\hbar}\,\textrm{Im}\Bigg\{ \\
 &\sum_{\beta\delta}V^{D,\alpha,\beta,\delta}_{\bq,-\bl}
\frac{1}{E_{\alpha,\bq}-E_{\beta,\bq-\bl} - E_{\delta,\bl} -i\Gamma }\times\\
\Big\{&(V^{D,\alpha,\beta,\delta}_{\bq,-\bl}-V^{X,\alpha,\beta,\delta}_{\bq,-\bl})^*
 (n_{\beta,\bq-\bl}n_{\delta,\bl} -n_{\alpha,\bq}( n_{\delta,\bl}+1)) \\
 + &(V^{D,\alpha,\delta,\beta}_{\bq,-\bq+\bl}-V^{X,\alpha,\delta,\beta}_{\bq,-\bq+\bl})^*
 (n_{\beta,\bq-\bl} n_{\delta,\bl} - n_{\alpha,\bq}(n_{\beta,\bq-\bl} +1))
 \Big\}
 \\
 -&\sum_{\beta\delta}    V^{D,\beta,\alpha,\delta}_{\bq+\bl,-\bl}       
 \frac{1}{E_{\beta,\bq+\bl}-E_{\alpha,\bq} - E_{\delta,\bl} -i\Gamma }\times\\
\Big\{&(V^{D,\beta,\alpha,\delta}_{\bq+\bl,-\bl} - V^{X,\beta,\alpha,\delta}_{\bq+\bl,-\bl})^*   
 (n_{\alpha,\bq}n_{\delta,\bl} -n_{\beta,\bq+\bl}( n_{\delta,\bl}+1)) \\+& (V^{D,\beta,\delta,\alpha}_{\bq+\bl,-\bq}-V^{X,\beta,\delta,\alpha}_{\bq+\bl,-\bq})^*
 (n_{\alpha,\bq} n_{\delta,\bl}-n_{\beta,\bq+\bl}(n_{\alpha,\bq}+1))
 \Big\}
 \\
 -&\sum_{\beta\delta}    V^{X,\beta,\alpha,\delta}_{\bq+\bl,-\bl}       
 \frac{1}{E_{\beta,\bq+\bl}-E_{\alpha,\bq} - E_{\delta,\bl} -i\Gamma }\times\\
\Big\{&(V^{X,\beta,\alpha,\delta}_{\bq+\bl,-\bl}-V^{D,\beta,\alpha,\delta}_{\bq+\bl,-\bl})^*   
 (n_{\alpha,\bq}n_{\delta,\bl} -n_{\beta,\bq+\bl}( n_{\delta,\bl}+1)) \\+& (V^{X,\beta,\delta,\alpha}_{\bq+\bl,-\bq}-V^{D,\beta,\delta,\alpha}_{\bq+\bl,-\bq})^*
 (n_{\alpha,\bq} n_{\delta,\bl}-n_{\beta,\bq+\bl}(n_{\alpha,\bq}+1))
 \Big\}\Bigg\}
 \,.
\end{split}
\label{eq:eom_final}
\end{equation}
There are three scattering channels for the exciton in state $\big|\alpha,\boldsymbol{q}\big>$ corresponding to the different roles that the exciton can play in an 
EEA process, see Fig.~\ref{fig:three_channels}.
Each channel contains a sum of two terms corresponding to the exhange of a whole exciton, e.g. states $\ket{\beta,\bq-\bl}$ and $\ket{\delta,\bl}$ for the first channel, 
due to the bosonic symmetry of triplets. The bosonic symmetry is combined with fermionic symmetries corresponding to the exchange of an electron or hole between two excitons, which becomes visible in the Bloch state representation of triplets, e.g. in Eq.~(\ref{eq:three_particle_bloch}).
Note that the interpretation of terms is complicated by these quantum symmetries, since particles can change roles in the scattering process.
In particular, the EOM do not have the form of a Boltzmann equation with terms like 
$V^{D,\alpha,\beta,\delta}_{\bq,-\bl}(V^{D,\alpha,\beta,\delta}_{\bq,-\bl}-V^{X,\alpha,\beta,\delta}_{\bq,-\bl})^*((1+n_{\alpha,\bq})n_{\beta,\bq-\bl}n_{\delta,\bl}-n_{\alpha,\bq}(1+n_{\beta,\bq-\bl})(1+n_{\delta,\bl})) $.
Even though we have formulated our theory in terms of exciton or two-particle quantities, it has been derived from a fundamental electron-electron Coulomb interaction. 
Hence it knows about the compound nature of excitons as bound electron-hole pairs.
In this context, we compare the fully microscopic theory to a purely bosonic effective theory in the next section.

\subsection{Bosonic EEA theory}

In the following, we set up an excitonic Hamiltonian to investigate if EEA can be understood in an effective bosonic picture. The Hamiltonian that describes
the process sketched in Fig.~\ref{fig:three_channels} is:
\begin{equation}
 \begin{split}
 H=\sum_{\alpha,\beta,\delta,\bq,\bl}\Big(V^{\alpha,\beta,\delta}_{\bq,\bl}  &X^{\dagger}_{\delta,-\bl}X^{\dagger}_{\beta,\bq+\bl}X^{\phantom\dagger}_{\alpha,\bq} \\
        +(V^{\alpha,\beta,\delta}_{\bq,\bl})^*  &X^{\dagger}_{\alpha,\bq}X^{\phantom\dagger}_{\beta,\bq+\bl}X^{\phantom\dagger}_{\delta,-\bl}\Big)
 \,,
\end{split}
\label{eq:bose_Hamilton}
\end{equation}
with effective exciton-exciton interaction matrix elements $V^{\alpha,\beta,\delta}_{\bq,\bl}$ and bosonic operators that fulfill 
$[X^{\phantom\dagger}_{\alpha,\bq},X^{\dagger}_{\beta,\bq'} ]=\delta_{\alpha\beta}\delta_{\bq,\bq'}$. As in the previous section,
we use the cluster expansion technique to obtain a closed set of dynamical equations. First, we derive EOM for triplets (without the trivial oscillating part),
where we discard all factorizations that contain non-density-like correlation functions:
\begin{equation}
\begin{split}
&i\hbar\frac{d}{dt}\Delta\left<X^{\dagger}_{\alpha,\bq_1}X^{\phantom\dagger}_{\beta,\bq_2}X^{\phantom\dagger}_{\delta,-\bl}  \right>\Big|_{\textrm{EEA}} \\
= &\Big(V^{\alpha,\beta,\delta}_{\bq_1,\bl}+V^{\alpha,\delta,\beta}_{\bq_1,-\bq_2}\Big)\times \\
& \Big\{n_{\alpha,\bq_1} n_{\delta,-\bl}-n_{\beta,\bq_2} n_{\delta,-\bl}+n_{\alpha,\bq_1} (1+n_{\beta,\bq_2})\Big\} \\
= &\Big(V^{\alpha,\beta,\delta}_{\bq_1,\bl}+V^{\alpha,\delta,\beta}_{\bq_1,-\bq_2}\Big)\times \\
& \Big\{n_{\alpha,\bq_1}(1+ n_{\delta,-\bl})(1+n_{\beta,\bq_2})-(1+n_{\alpha,\bq_1})n_{\delta,-\bl}n_{\beta,\bq_2}\Big\}
\,.
\end{split}
\label{eq:triplet_final_boson}
\end{equation}
For the exciton densities we obtain:
\begin{equation}
\begin{split}
 \frac{d}{dt}n_{\alpha,\bq}&= 
 \sum_{\beta,\delta,\bl}
 \times \\
  \frac{2}{\hbar}\,\textrm{Im}\Bigg\{ \Big(V^{\alpha,\beta,\delta}_{\bq, \bl}\Big)^*&\Delta\left<X^{\dagger}_{\alpha,\bq}X^{\phantom\dagger}_{\beta,\bq+\bl}X^{\phantom\dagger}_{\delta,-\bl} \right> \\
 +\Big(V^{\beta,\alpha,\delta}_{\bq+\bl,-\bl}+V^{\beta,\delta,\alpha}_{\bq+\bl,-\bq}\Big)
 &\Big(\Delta\left<X^{\dagger}_{\beta,\bq+\bl}X^{\phantom\dagger}_{\delta,\bl}X^{\phantom\dagger}_{\alpha,\bq} \right>\Big)^*
 \Bigg\}\,.
\end{split}
\label{eq:EOM_n_X_boson}
\end{equation}
Inserting the triplets in Markov approximation yields:
\begin{equation}
\begin{split}
 &\frac{d}{dt}n_{\alpha,\bq}= 
 \sum_{\beta,\delta,\bl}\frac{2}{\hbar}\,\textrm{Im}\Bigg\{
\\
   &V^{\alpha,\beta,\delta}_{\bq, -\bl}
  \Big(V^{\alpha,\beta,\delta}_{\bq,-\bl}+V^{\alpha,\delta,\beta}_{\bq,-\bq+\bl}\Big)^*\frac{1}{E_{\alpha,\bq}-E_{\beta,\bq-\bl} - E_{\delta,\bl} -i\Gamma }\times \\
& \Big\{(1+n_{\alpha,\bq})n_{\delta,\bl}n_{\beta,\bq-\bl}-n_{\alpha,\bq}(1+ n_{\delta,\bl})(1+n_{\beta,\bq-\bl})\Big\} \\
 -&V^{\beta,\alpha,\delta}_{\bq+\bl,-\bl}
 \Big(V^{\beta,\alpha,\delta}_{\bq+\bl,-\bl}+V^{\beta,\delta,\alpha}_{\bq+\bl,-\bq}\Big)^*\frac{1}{E_{\beta,\bq+\bl}- E_{\alpha,\bq}- E_{\delta,\bl} -i\Gamma }\times \\
 & \Big\{(1+n_{\beta,\bq+\bl})n_{\delta,\bl}n_{\alpha,\bq}-n_{\beta,\bq+\bl}(1+ n_{\delta,\bl})(1+n_{\alpha,\bq})\Big\} \\
 -&V^{\beta,\delta,\alpha}_{\bq+\bl,-\bq}
 \Big(V^{\beta,\delta,\alpha}_{\bq+\bl,-\bq}+V^{\beta,\alpha,\delta}_{\bq+\bl,-\bl}\Big)^*\frac{1}{E_{\beta,\bq+\bl}- E_{\alpha,\bq}- E_{\delta,\bl} -i\Gamma }\times \\
 & \Big\{(1+n_{\beta,\bq+\bl})n_{\delta,\bl}n_{\alpha,\bq}-n_{\beta,\bq+\bl}(1+ n_{\delta,\bl})(1+n_{\alpha,\bq})\Big\}
 \Bigg\}\,.
\end{split}
\label{eq:EOM_n_X_final_boson}
\end{equation}
The scattering integrals on the RHS have to be compared to those in Eq.~(\ref{eq:eom_final}) derived from the fundamental electron-electron interaction Hamiltonian.
Both equations exhibit three main scattering channels, with two terms for each channel reflecting the indistinguishablity of the scattering excitons. 
In Eq.~(\ref{eq:EOM_n_X_final_boson}), the two terms are given by sums over matrix elements such as $V^{\alpha,\beta,\delta}_{\bq,-\bl}+V^{\alpha,\delta,\beta}_{\bq,-\bq+\bl}$.
Unlike the fully microscopic EOM, the bosonic EOM have the form of a Boltzmann equation in the sense that population factors $n$ and $1+n$ can be clearly assigned to 
scattering out of an exciton state and scattering into an exciton state, respectively. The assignment of in- and out-scattering is consistent with the (approximate) 
energy conservation as well as the exciton-exciton interaction matrix elements, where $V^{\alpha,\beta,\delta}_{\bq,-\bl}$ belongs to scattering between
$\ket{\alpha,\bq}$ and $\ket{\beta,\bq+\bl}$, $\ket{\delta,-\bl}$. The key difference is that the purely bosonic theory can not account for the exchange of a single 
electron or hole between two excitons (reflected by a change from $V^{D,\alpha,\beta,\delta}_{\bq,-\bl} $ to $-V^{X,\alpha,\beta,\delta}_{\bq,-\bl} $), but only of an exciton as a whole ($V^{\alpha,\beta,\delta}_{\bq,-\bl}$ to $V^{\alpha,\delta,\beta}_{\bq,-\bq+\bl} $). 
To clarify this, we compare the EOM for triplets as derived from the two Hamiltonians term by term:
\begin{equation}
\begin{split}
&i\hbar\frac{d}{dt}\Delta\left<X^{\dagger}_{\alpha,\bq_1}X^{\phantom\dagger}_{\beta,\bq_2}X^{\phantom\dagger}_{\delta,-\bl}  \right>\Big|_{\textrm{bosonic}} \\
= &V^{\alpha,\beta,\delta}_{\bq_1,\bl}\Big\{\underbrace{n_{\alpha,\bq_1} n_{\delta,-\bl}}_{\#1}-\underbrace{n_{\beta,\bq_2} n_{\delta,-\bl}}_{\#2}+\underbrace{n_{\alpha,\bq_1}}_{\#3}+\underbrace{n_{\alpha,\bq_1}n_{\beta,\bq_2}}_{\#4}\Big\} \\
+&V^{\alpha,\delta,\beta}_{\bq_1,-\bq_2}\Big\{\underbrace{n_{\alpha,\bq_1} n_{\delta,-\bl}}_{\#5}-\underbrace{n_{\beta,\bq_2} n_{\delta,-\bl}}_{\#6}+\underbrace{n_{\alpha,\bq_1}}_{\#7}+\underbrace{n_{\alpha,\bq_1}n_{\beta,\bq_2}}_{\#8}\Big\}
\,, \\
&i\hbar\frac{d}{dt}\Delta\left<X^{\dagger}_{\alpha,\bq_1}X^{\phantom\dagger}_{\beta,\bq_2}X^{\phantom\dagger}_{\delta,-\bl}  \right>\Big|_{\textrm{full}} \\
= &V^{D,\alpha,\beta,\delta}_{\bq_1,\bl}\underbrace{n_{\alpha,\bq_1} n_{\delta,-\bl}}_{\#1}-\Big(V^{D,\alpha,\beta,\delta}_{\bq_1,\bl}-V^{X,\alpha,\beta,\delta}_{\bq_1,\bl}\Big)\underbrace{n_{\beta,\bq_2} n_{\delta,-\bl}}_{\#2}\\
+&\Big(V^{D,\alpha,\beta,\delta}_{\bq_1,\bl}-V^{X,\alpha,\beta,\delta}_{\bq_1,\bl}\Big)\underbrace{n_{\alpha,\bq_1}}_{\#3}-V^{X,\alpha,\delta,\beta}_{\bq_1,-\bq_2}\underbrace{n_{\alpha,\bq_1}n_{\beta,\bq_2}}_{\#4} \\
-&V^{X,\alpha,\beta,\delta}_{\bq_1,\bl}\underbrace{n_{\alpha,\bq_1} n_{\delta,-\bl}}_{\#5}-\Big(V^{D,\alpha,\delta,\beta}_{\bq_1,-\bq_2}-V^{X,\alpha,\delta,\beta}_{\bq_1,-\bq_2}\Big)\underbrace{n_{\beta,\bq_2} n_{\delta,-\bl}}_{\#6}\\
+&\Big(V^{D,\alpha,\delta,\beta}_{\bq_1,-\bq_2}-V^{X,\alpha,\delta,\beta}_{\bq_1,-\bq_2}\Big)\underbrace{n_{\alpha,\bq_1}}_{\#7}+V^{D,\alpha,\delta,\beta}_{\bq_1,-\bq_2}\underbrace{n_{\alpha,\bq_1}n_{\beta,\bq_2}}_{\#8}\,.
\end{split}
\label{eq:triplet_comparison}
\end{equation}
While in the purely bosonic picture, terms can be collected such that Boltzmann-like population factors emerge, this is not possible in the full theory due to the exchange matrix elements.
The
possible exchange of fermionic constituents of excitons is what inhibits the
fully microscopic EOM to be cast into the form of a Boltzmann equation with effective exciton-exciton interaction matrix elements.
This is consistent with the more general discussion by M. Combescot et al. that it is not possible to formulate a closed expression for an 
effective exciton-exciton interaction potential \cite{combescot_effective_2002}.
\\We finally compare the two pictures in the limiting case where the exciton distribution is close to equilibrium due to fast relaxation processes.
Focussing on the EOM of 1s-exciton populations and neglecting the populations of high-energy states, we find:
\begin{equation}
 \begin{split}
\frac{d}{dt} n_{\alpha,\bq}&\Big|_{\textrm{full}}= 
 -\frac{1}{\mathcal{A}^2}\sum_{\bl}\sum_{\beta\delta}
 \frac{2}{\hbar}\times \\
 \textrm{Im}\Bigg\{&   (V^{D,\beta,\alpha,\delta}_{\bq+\bl,-\bl}-V^{X,\beta,\alpha,\delta}_{\bq+\bl,-\bl})\times       
 \\
\Big(&(V^{D,\beta,\alpha,\delta}_{\bq+\bl,-\bl} - V^{X,\beta,\alpha,\delta}_{\bq+\bl,-\bl})^*+ (V^{D,\beta,\delta,\alpha}_{\bq+\bl,-\bq}-V^{X,\beta,\delta,\alpha}_{\bq+\bl,-\bq})^*\Big)\times\\
&\frac{1}{E_{\beta,\bq+\bl}-E_{\alpha,\bq} - E_{\delta,\bl} -i\Gamma }
 \Bigg\}n_{\alpha,\bq}n_{\delta,\bl}
\end{split}
\label{eq:eom_simplified}
\end{equation}
and
\begin{equation}
 \begin{split}
\frac{d}{dt} n_{\alpha,\bq}&\Big|_{\textrm{bosonic}}= 
 -\frac{1}{\mathcal{A}^2}\sum_{\bl}\sum_{\beta\delta}
 \frac{2}{\hbar}\Big|V^{\beta,\alpha,\delta}_{\bq+\bl,-\bl}+V^{\beta,\delta,\alpha}_{\bq+\bl,-\bq}\Big|^2\times \\
&\textrm{Im}\Bigg\{\frac{1}{E_{\beta,\bq+\bl}-E_{\alpha,\bq} - E_{\delta,\bl} -i\Gamma }
 \Bigg\}n_{\alpha,\bq}n_{\delta,\bl}\,.
\end{split}
\label{eq:eom_simplified_boson}
\end{equation}
While the effective bosonic picture allows to identify the modulus square of an exciton-exciton interaction matrix element that obeys bosonic symmetry, this is still not possible 
in the fully microscopic theory.
\begin{figure}
\centering
\includegraphics[width=\columnwidth]{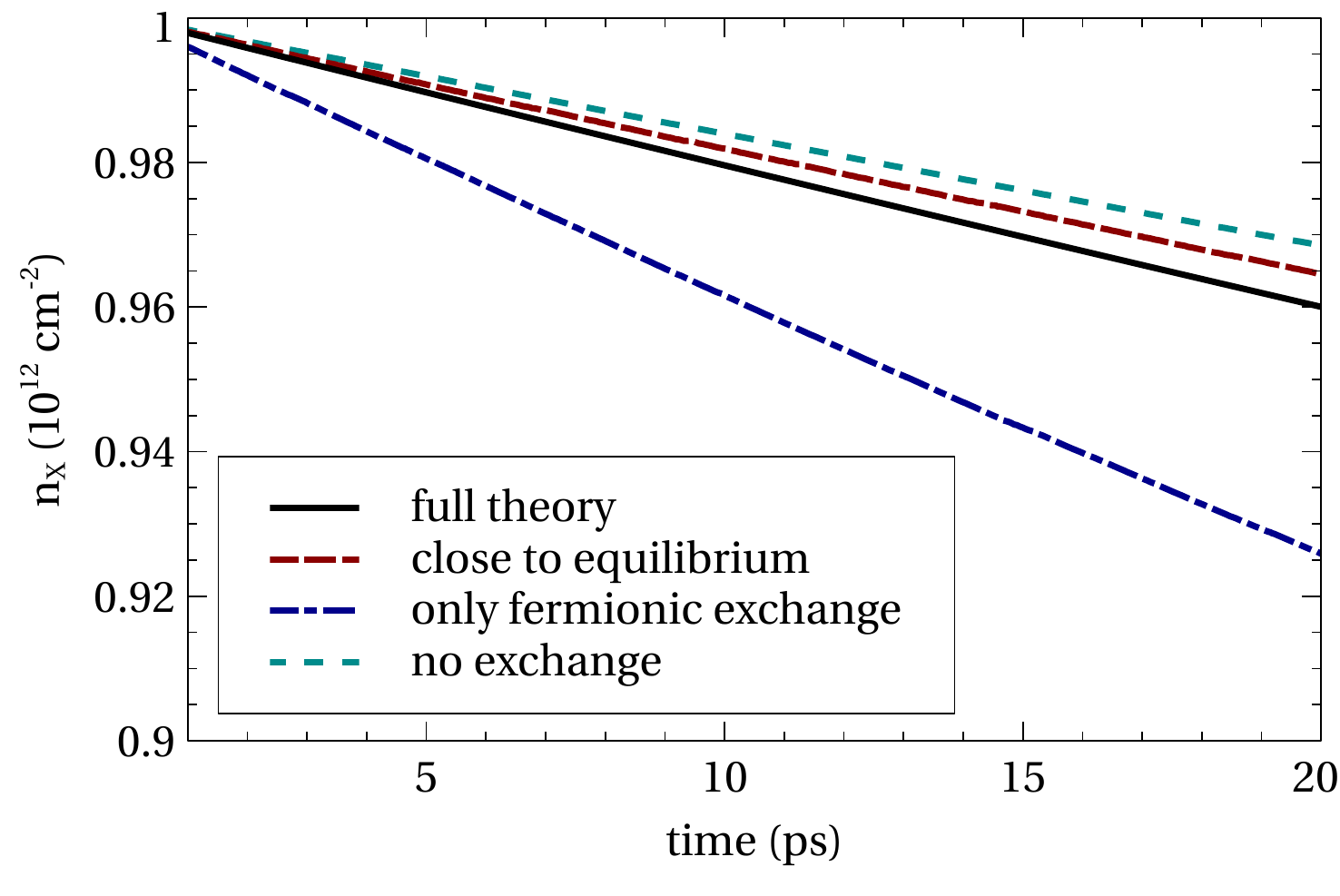}
\caption{Decay of the total exciton density in hBN-encapsulated MoS$_2$ at $T=300$ K obtained from different approximations to our EEA theory. We compare the full theory (Eq.~(\ref{eq:eom_final})) to the simplified EOM in the close-to-equilibrium limit (Eq.~(\ref{eq:eom_simplified})). Further approximations that lead to an equivalence of Eqs.~(\ref{eq:eom_simplified}) and (\ref{eq:eom_simplified_boson}) are performed by including only fermionic exchange and by neglecting all exchange effects. The corresponding EEA coefficients are given in the text.
}
\label{fig:comp}
\end{figure}
\\An equivalence between the full theory and the bosonic theory can only be obtained when the exchange of whole excitons is neglected, which amounts to discarding the Coulomb matrix elements $V^{\beta,\delta,\alpha}_{\bq+\bl,-\bq}$ in Eqs.~(\ref{eq:eom_simplified}) and (\ref{eq:eom_simplified_boson}). Then the effective bosonic interaction matrix element can be identified as $V^{D,\beta,\alpha,\delta}_{\bq+\bl,-\bl}-V^{X,\beta,\alpha,\delta}_{\bq+\bl,-\bl}$, which means that fermionic exchange can still be included in this case. In Fig.~\ref{fig:comp}, we demonstrate the quantitative effect of this approximation. First of all, when reducing the full theory to the close-to-equilibrium case, the EEA coefficient decreases from $0.21\times10^{-3}$ cm$^{2}$s$^{-1}$ to $0.18\times10^{-3}$ cm$^{2}$s$^{-1}$. Neglecting the exchange of full excitons leads to an increase of the coefficient to $0.40\times10^{-3}$ cm$^{2}$s$^{-1}$. The coefficient is reduced again to $0.16\times10^{-3}$ cm$^{2}$s$^{-1}$ in the absence of all exchange processes, which means that EEA is even slower than in the full theory. We conclude that an effective bosonic theory that includes fermionic exchange effects overestimates EEA efficiency in encapsulated MoS$_2$ by a factor $2$. This is partly remedied by neglecting fermionic exchange as well due to a compensation between the different exchange effects.

\subsection{Brillouin zone sampling and convergence}

\begin{figure}
\centering
\includegraphics[width=\columnwidth]{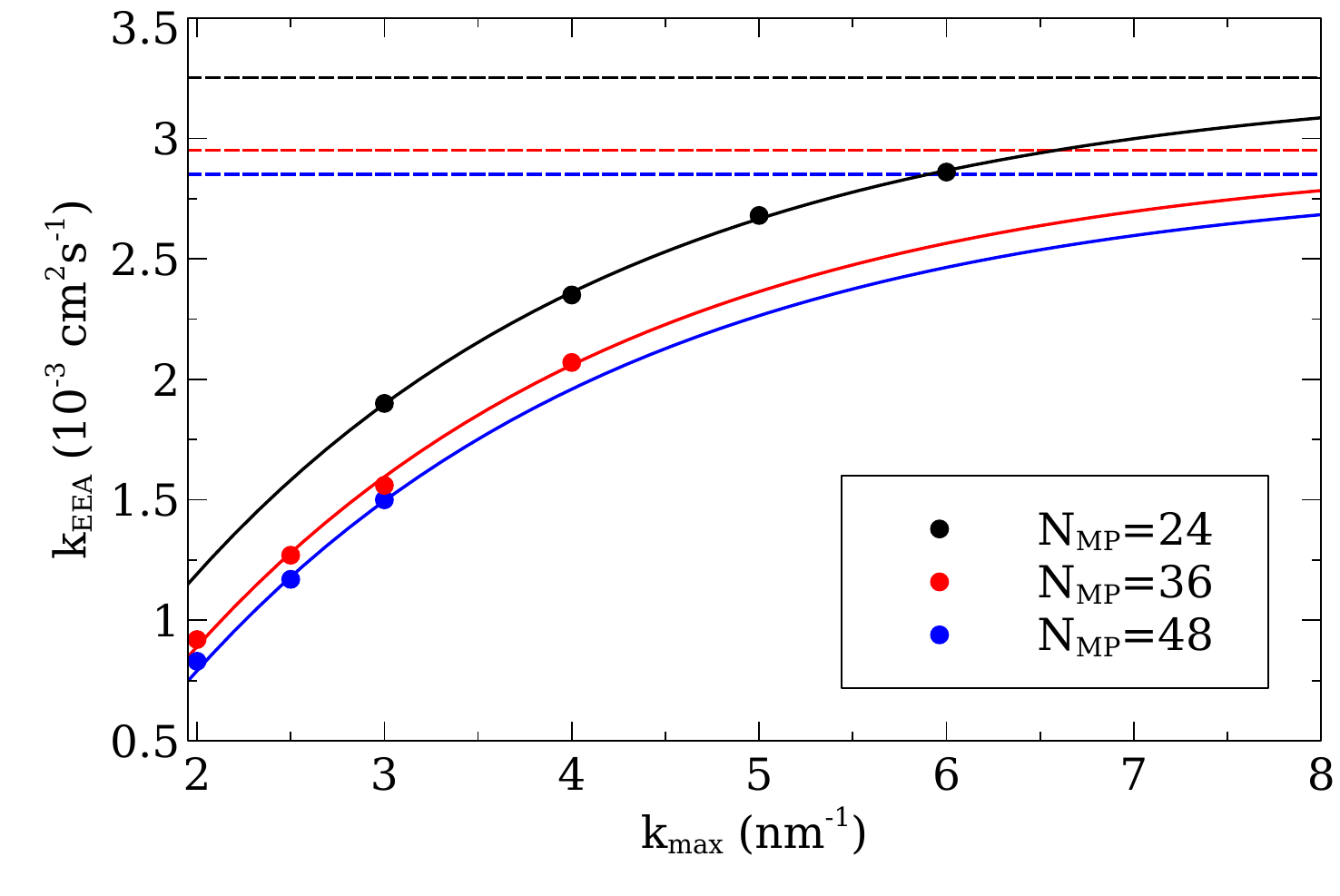}
\caption{Dependence of EEA coefficient in hBN-encapsulated MoS$_2$ at $T=300$ K on the Brillouin-zone sampling. We compare Monkhorst-Pack grids with $N_{\textrm{MP}}\times N_{\textrm{MP}}\times 1$ grid points for different radii $k_{\textrm{max}}$ of the region taken into account around the K-point. The numerical results are shown as filled circles, while analytic fitting curves of the form $f(x)=a-\textrm{exp}(-b\cdot(x-c))$ are represented by solid lines. The dashed lines correspond to the asymptotic EEA coefficient obtained from the analytic functions for large radii.
}
\label{fig:k_conv}
\end{figure}

The numerical simulation of EEA involves several steps: the diagonalization of the BSE (2) to obtain two-particle energies and wave functions, the calculation of exciton-exciton interaction matrix elements (5) and the propagation of the EOM for exciton populations (4) and (7). To this end, an appropriate sampling of the Brillouin zone has to be applied, where the number of grid points is constrained by the high-dimensionality of the problem. As explained in the manuscript, we focus on Bloch states in the K-valley, using a Monkhorst-Pack grid to sample the Brillouin zone in a circle with radius $k_{\textrm{max}}$ around the K-point. Most results are based on a $36\times36\times1$-grid and $k_{\textrm{max}}=4$ nm$^{-1}$, propagating the EOM until $t_{\textrm{max}}=20$ ps to extract EEA coefficients $k_{\textrm{EEA}}$ via the analytic fit formula 
$n_{\textrm{X}}(t)= n_{\textrm{X},0}(1+n_{\textrm{X},0}k_{\textrm{EEA}}t)^{-1}$. One exception is the time dependence of the total exciton density $n_{\textrm{X}}(t)$ shown in Fig.2(b) for illustrative purposes. Here, we used a $24\times24\times1$-grid and $k_{\textrm{max}}=3$ nm$^{-1}$ to reduce the numerical effort for propagating the EOM until $t_{\textrm{max}}=500$ ps.
\\The convergence of results with respect to Brillouin zone sampling is shown in Fig.~\ref{fig:k_conv} for hBN-encapsulated MoS$_2$ at $T=300$ K. The convergence with respect to $k_{\textrm{max}}$ is slow, but can be well extrapolated by analytic fitting curves with exponential asymptotics. Note that the distance between K and K' is $13.3$ nm$^{-1}$ for the given lattice constant. We find that results are almost converged with respect to k-point density for a $36\times36\times1$-grid. From the analysis of asymptotics, we estimate a converged value $k_{\textrm{EEA}}=2.8\times10^{-3}$ cm$^{2}$s$^{-1}$, which is about $35\%$ larger than the value obtained with our standard grid.

\begin{figure}
\centering
\includegraphics[width=\columnwidth]{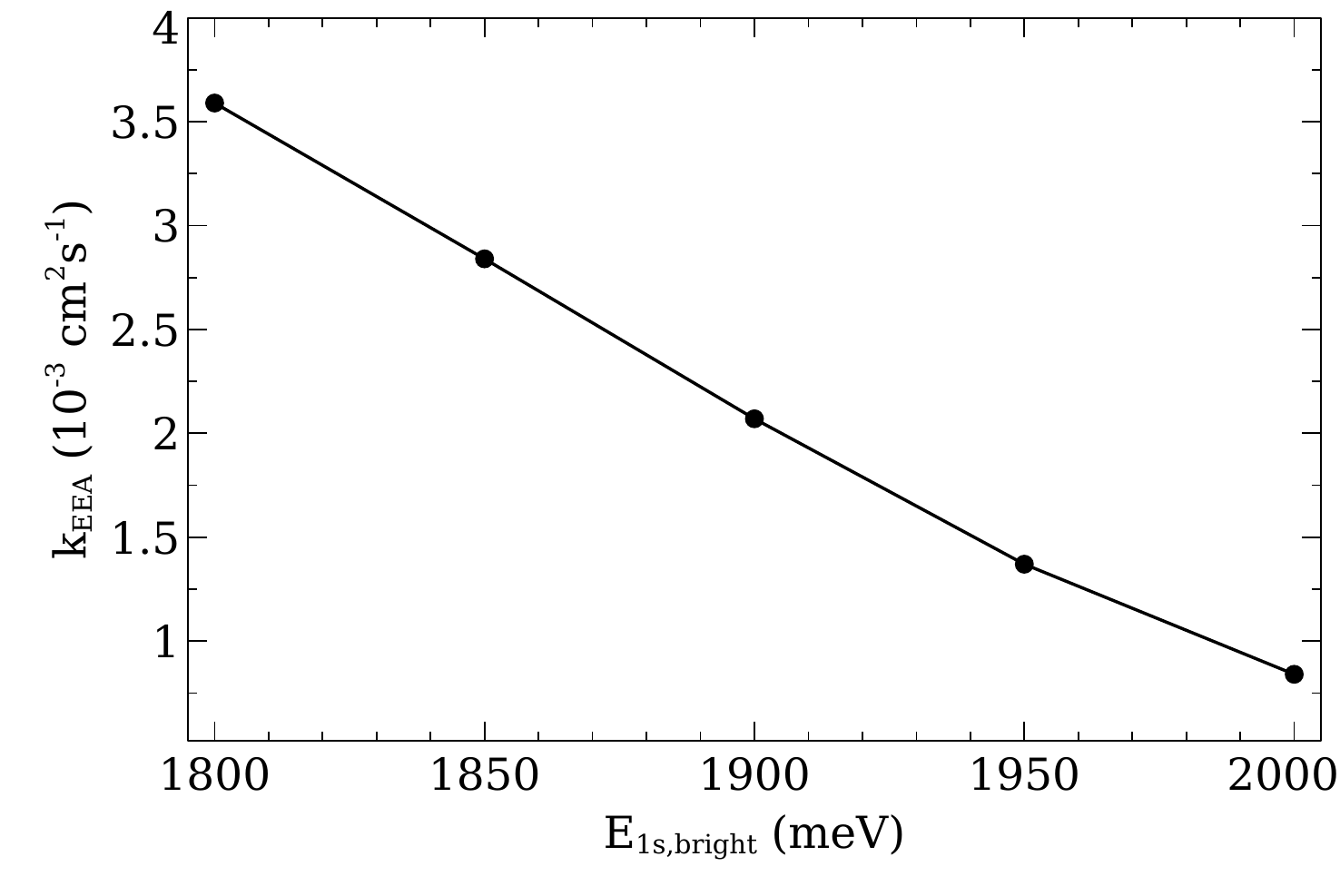}
\caption{Dependence of EEA coefficient in hBN-encapsulated MoS$_2$ at $T=300$ K on the energy of the exciton ground state $E_{\textrm{1s,bright}}$, which corresponds to 
the like-spin exciton at the K-point with zero total momentum (so-called A-exciton).}
\label{fig:E_X_dep}
\end{figure}

\subsection{Influence of 1s-exciton energy}

The dependence of the EEA coefficient on the ground-state exciton energy $E_{\textrm{1s,bright}}$ is shown in Fig.~\ref{fig:E_X_dep}. In the manuscript, $E_{\textrm{1s,bright}}=1900$ meV is used. An increase (decrease) of $E_{\textrm{1s,bright}}$ corresponds to an increase (decrease) of the average energy of high-energy target states for the EEA process, see Fig.~2(a) and (c). As a trend, we find that EEA becomes more efficient for smaller $E_{\textrm{1s,bright}}$. Since in this situation the target states move to lower energies, it is equivalent to a hypothetical movement of the third conduction band upwards relative to the first conduction band, which could be due to uncertainties in the underlying first-principle calculation.

\begin{figure}
\centering
\includegraphics[width=\columnwidth]{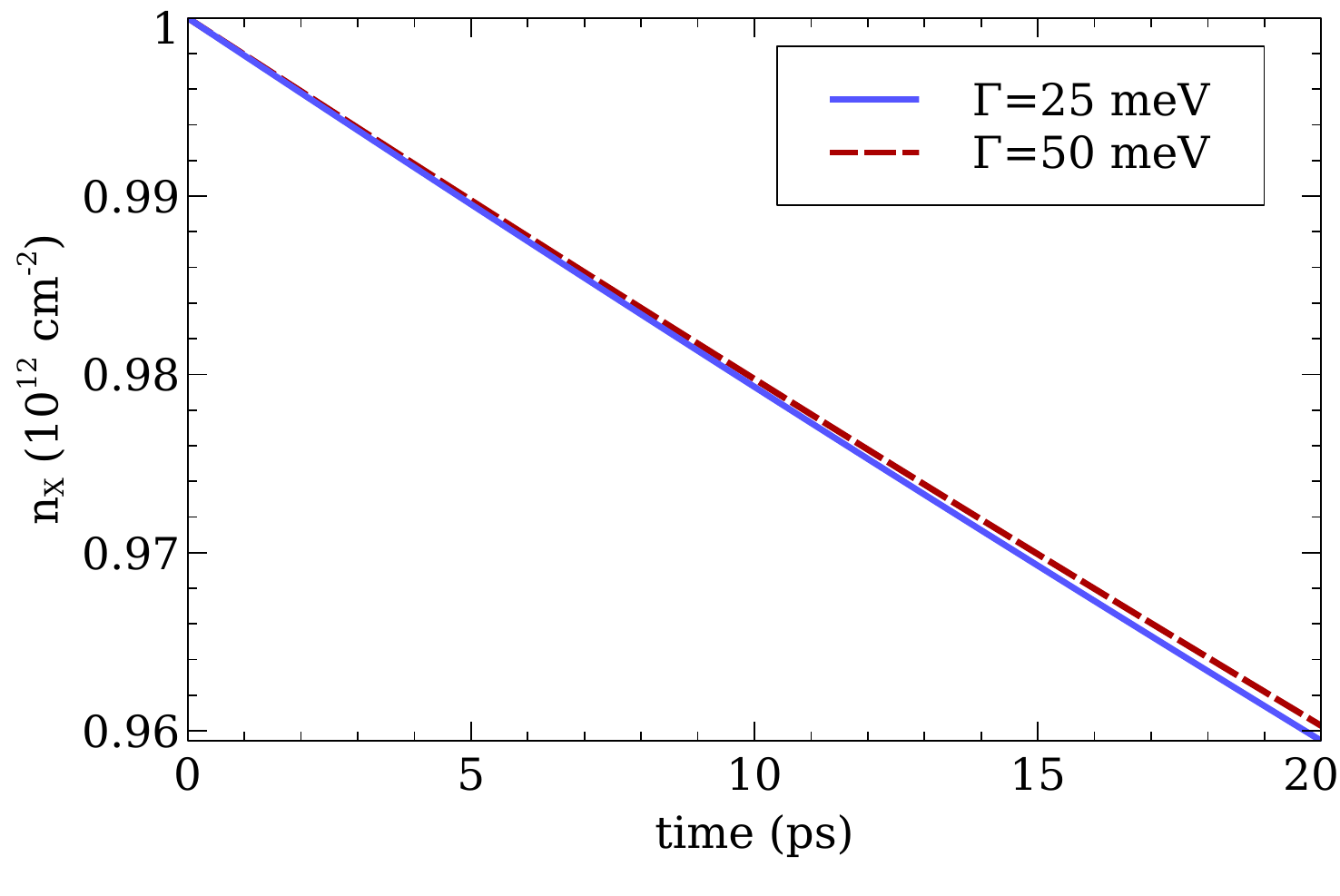}
\caption{ Time dependence of the total exciton density $n_{\textrm{X}}$ in hBN-encapsulated MoS$_2$ at $T=300$ K for two different phenomenological damping constants $\Gamma$, 
see Eq.~(4) in the manuscript. The EEA coefficients according to Eq.~(8) are $k_{\textrm{EEA}}=2.07\times10^{-3}$ cm$^{2}$s$^{-1}$ for $\Gamma=50$ meV and
$k_{\textrm{EEA}}=2.11\times10^{-3}$ cm$^{2}$s$^{-1}$ for $\Gamma=25$ meV.}
\label{fig:Gamma_dep}
\end{figure}

\subsection{Influence of phenomenological damping}

We compare the time evolution of the total exciton density $n_{\textrm{X}}$ for the phenomenological damping used in the manuscript ($\Gamma=50$ meV) with a calculation
using $\Gamma=25$ meV in Fig.~\ref{fig:Gamma_dep}. The Brillouin zone is sampled with our standard k-mesh. We find a weak dependence on $\Gamma$ in the range of several percent.

\end{document}